\documentclass[13 pt]{article}
\usepackage{amssymb}

\newcommand{\finproof}{{\hfill \rule{5pt}{5pt}\\}}
\newtheorem{definition}{Definition}[section]

\newtheorem{lemma}[definition]{Lemma}
\newtheorem{corollary}[definition]{Corollary}
\newtheorem{remark}[definition]{Remark}

\newtheorem{proposition}[definition]{Proposition}
\def\acal{{\Large{\textit{a}}}}
\def\bcal{{\Large{\textit{b}}}}
\def\ccal{{\Large{\textit{c}}}}
\def\xcal{{\Large{\textit{x}}}}
\def\acalt{\tilde{\Large{\textit{a}}}}
\def\bcalt{\tilde{\Large{\textit{b}}}}
\def\ccalt{\tilde{\Large{{\textit{c}}}}}
\def\binombN{\left(^{N}_{n+1}\right)}
\def\binomaN{\left(^{N}_{n}\right)}
\def\binomcN{\left(^{N}_{n-1}\right)}
\def\binomNmun{\left(^{N-1}_{n}\right)}
\def\binomN{\left(^{N}_{n}\right)}
\def\binomNmunpun{\left(^{N-1}_{n+1}\right)}
\def\binomNpun{\left(^{N}_{n+1}\right)}
\def\binomNmunmun{\left(^{N-1}_{n-1}\right)}
\def\binomNnmun{\left(^{N}_{n-1}\right)}
\def\binomNnmdeux{\left(^{N}_{n-2}\right)}
\def\binomNs{\left(^{N}_{s}\right)}
\def\binomNspun{\left(^{N}_{s+1}\right)}
\def\binomNsmun{\left(^{N}_{s-1}\right)}

\def\lambdat{\tilde{\lambda}}

\newcommand{\beq}{\begin{equation}}
\newcommand{\eeq}{\end{equation}}
\newcommand{\bea}{\begin{eqnarray*}}
\newcommand{\eea}{\end{eqnarray*}}
\newcommand{\beqa}{\begin{eqnarray}}
\newcommand{\eeqa}{\end{eqnarray}}

\begin{document}

\newfont{\elevenmib}{cmmib10 scaled\magstep1}%

\newcommand{\Title}[1]{{\baselineskip=26pt \begin{center}
            \Large   \bf #1 \\ \ \\ \end{center}}}
\hspace*{2.13cm}%
\hspace*{1cm}%
\newcommand{\Author}{\begin{center}\large
           Pascal Baseilhac\footnote{
baseilha@phys.univ-tours.fr} 
\end{center}}
\newcommand{\Address}{{\baselineskip=18pt \begin{center}
           \it Laboratoire de Math\'ematiques et Physique Th\'eorique CNRS/UMR 6083,\\
           F\'ed\'eration Denis Poisson,\\
Universit\'e de Tours, Parc de Grandmont, 37200 Tours, France
      \end{center}}}
\baselineskip=13pt

\bigskip
\vspace{-1cm}

\Title{A family of tridiagonal pairs and related symmetric functions}\Author

\vspace{- 0.1mm}
 \Address

\vskip 0.6cm

\centerline{\bf Abstract}\vspace{0.3mm}  \vspace{1mm}
A family of tridiagonal pairs which appear in the context of quantum integrable systems is studied in details. The corresponding eigenvalue sequences, eigenspaces and the block tridiagonal structure of their matrix realizations with respect the dual eigenbasis are described. The overlap functions between the two dual basis are shown to satisfy a coupled system of recurrence relations and a set of discrete second-order $q-$difference equations which generalize the ones associated with the Askey-Wilson orthogonal polynomials with a discrete argument. Normalizing the fundamental solution to unity, the hierarchy of solutions are rational functions of one discrete argument, explicitly derived in some simplest examples. The weight function which ensures the orthogonality of the system of rational functions defined on a discrete real support is given.
\vspace{0.2cm} 

{\small 2000 MSC:\ 20G42;\ 39A13;\ 42C05}



{{\small  {\it \bf Keywords}: Tridiagonal pair; Recurrence relations; $q-$difference equations; Orthogonal symmetric functions}}
%
%

\section{Introduction}
Jacobi matrices deserve continuous attention in operator theory, as they play an important role in various areas such as numerical analysis, orthogonal polynomials, continued fractions. They also find applications in mathematical physics: they are discrete analogues of second-order linear differential operators of Schr${\ddot{o}}$dinger type on the half-line, and appear in integrable systems and the theory of random matrices.\vspace{1mm} 

An important domain of application concerns the theory of orthogonal polynomials of one argument $x$ - denoted $p_n(x)$ below. Indeed, it is well-known that every sequence $\{p_n\}_{n=0}^{n=\infty}$ satisfies a three-term relation
\beqa
xp_n = b_n p_{n+1}+a_n p_{n}+c_n p_{n-1}\ , \quad n=1,2,...\label{rec0}
\eeqa
where $p_0\equiv 1$, $c_{0}\equiv 0$ by definition, and $a_n,b_n,c_n\in {\mathbb C}$. A Jacobi matrix $J$ being a tridiagonal matrix of either finite or infinite dimension, in the infinite dimensional case the non-vanishing coefficients $a_n,b_n,c_n$ yield to consider the spectral problem associated with the sequence $\{p_n\}_{n=0}^{n=\infty}$:
\beqa
J p_n(x)=x p_n(x) \qquad \mbox{with} \qquad \ J = \left(
\begin{array}{ c c c c c c}
a_0 & c_1  &      &      &   & 0\\
b_0 & a_1  &  c_2   &      &   &  \\
  & b_1  &  a_2    & \cdot  &   & \\
  &   &  \cdot    & \cdot  & \cdot   & \\
  &   &           &  \cdot & \cdot &  \\
  0 &   &   &   &  &   
\end{array}\label{spec}
\right),
\eeqa
where $p_{0}(x),p_{1}(x),p_{2}(x),...$ form a basis of ${\mathbb C}[x]$. In the finite dimensional case, the main submatrix of $J$ which is called the truncated matrix and denoted $J^{(N)}$ is usually introduced. If it is of size $N+1\times N+1$, then the zeros of $p_{N+1}(x)$ coincide with the $N+1$ distinct eigenvalues of $J^{(N)}$. In other words, $p_{N+1}(x)$ is proportional to the characteristic polynomial of $J^{(N)}$ and $x$ takes discrete values. As a consequence, the corresponding system of orthogonal polynomials is defined on a discrete support.   

Among the families of orthogonal polynomials in one variable, the only ones satisfying (\ref{spec}) and a second-order differential equation
are known to be the Jacobi, Hermite, Laguerre and Bessel polynomials, as shown in a theorem of Bochner \cite{Boch}. This led the authors of \cite{Haine} to derive a ``$q-$version'' of this theorem using  an operator identity of independent interest. For this family of orthogonal polynomials satisfying (\ref{spec}) and a second-order $q-$difference equation of the form 
\beqa
a(y)\big(p_{n}(s(qy))-p_{n}(s(y))\big)+b(y)\big(p_{n}(s(q^{-1}y))-p_{n}(s(y))\big) = \theta_n p_{n}(s(y))\qquad \mbox{for}\qquad n\geq 0\ ,\label{qBoch}
\eeqa
where the argument is either $x\equiv s(y)=(y+y^{-1})/2$ or $x\equiv s(y)=y$, the only solutions to (\ref{spec}) and (\ref{qBoch}) are given by the Askey-Wilson polynomials \cite{AW} or the big $q-$Jacobi polynomials, respectively \cite{Haine}.\vspace{1mm}

In view of the tridiagonal structure of the l.h.s. of above equations, it is natural to study the algebraic structure associated with a pair of Jabobi matrices having a spectral problem of the form (\ref{spec}), (\ref{qBoch}). In particular, there has been different ways to characterize the Askey-Wilson polynomials and related $q-$hypergeometric functions from this {\it algebraic} point of view. For instance, let us mention \cite{Koor} where finite dimensional representations of the quantum group $U_q(sl_2)$ arise, and more generally \cite{Zhed92} where finite dimensional representations of the Askey-Wilson algebra  $AW(3)$ are introduced and studied in details.\vspace{1mm}

In the last few years, a unified algebraic framework based on the concept of {\it Leonard pair} has been introduced in connection with the orthogonal polynomials of the Askey scheme.
Roughly speaking (details can be found in \cite{TerLP01}), a Leonard pair $A,A^*$ acting on a finite dimensional representation $V$ is such that $A,A^*$ are diagonalizable where the matrix representing $A^*$ (resp. $A$) is irreducible tridiagonal in the basis which diagonalizes $A$ (resp. $A^*$). In particular, given a Leonard pair there exists a sequence of scalars
$\beta,\gamma,\gamma^*,\varrho,\varrho^*,
\omega,\eta,\eta^*$ taken from an arbitrary field $\mathbb K$ such that \cite{TerAW03}
\begin{eqnarray}
A^2 A^*-\beta A A^*\!A+A^*\!A^2-\gamma\left( A A^*\!+\!A^*\!A
\right)-\varrho\,A^* &=& \gamma^*\!A^2+\omega A+\eta\,I\ ,\nonumber\\
A^*{}^2\!A-\beta A^*\!AA^*\!+AA^*{}^2-\gamma^*\!\left(A^*\!A\!+\!A
A^*\right)-\varrho^*\!A &=&\gamma A^*{}^2+\omega
A^*\!+\eta^*\!I\ .\label{AWrel}
\end{eqnarray}
The sequence is uniquely determined by the Leonard pair provided the dimension of $V$ is at least $4$, and the equations above are
called the {\em Askey-Wilson relations} (see also \cite{Zhed92}). 
Remarkably, all known examples of Leonard pairs for $\beta=q+q^{-1}$ and $q\neq -1$ are related with orthogonal polynomials of the Askey scheme \cite{KoeSwa} in the following sense: the entries of the transition matrix (sometimes called the overlap coefficients) relating the two ``dual'' basis which diagonalize $A,A^*$, respectively,  can be expressed in terms of one of the following orthogonal polynomials ($q-$hypergeometric functions): Racah (${{}_4F_3}$), Hahn and dual Hahn (${{}_3F_2}$), Krawtchouk (${{}_2F_1}$), $q$-Racah (${{}_4\phi_3}$), $q$-Hahn and dual $q$-Hahn (${{}_3\phi_2}$), $q$-Krawtchouk - classical, affine, quantum, dual - (${{}_2\phi_1}$).  \vspace{1mm}

Given the connection between Leonard pairs and the polynomials of the Askey-scheme, several features led to consider a more general object called a {\it Tridiagonal} (TD) pair \cite{Ter01}. From an algebraic point of view, tridiagonal pairs play an important role in the representation theory of the tridiagonal algebra \cite{Ter01,Ter03,Ter93}. This associative algebra $\mathbb{T}$ with unity consists of two generators - called standard generators - acting on a vector space $V$, say ${\textsf A}:V\rightarrow V$ and ${\textsf A}^*:V\rightarrow V$. In general, the defining relations of $\mathbb{T}$ depend on five scalars $\rho,\rho^*,\gamma,\gamma^*$ and $\beta$. In the following, we will focus on the {\it reduced} parameter sequence $\gamma=0,\gamma^*=0$ and $\beta=q+q^{-1}$ which exhibits all interesting properties that can be extended to more general parameter sequences. In this case, the so-called {\it tridiagonal relations} take the form
\beqa
[\textsf{A},[\textsf{A},[\textsf{A},\textsf{A}^*]_q]_{q^{-1}}]=\rho[\textsf{A},\textsf{A}^*]\
,\qquad
[\textsf{A}^*,[\textsf{A}^*,[\textsf{A}^*,\textsf{A}]_q]_{q^{-1}}]=\rho^*[\textsf{A}^*,\textsf{A}]\
\label{qDG} , \eeqa
where the $q$-commutator $\
[\textsf{A},\textsf{B}]_q=q^{1/2}\textsf{A}\textsf{B}-q^{-1/2}\textsf{B}\textsf{A}$ has been introduced. Here, $q$ is a deformation parameter assumed to be not a root of
unity. Note that for $\rho=\rho^*=0$ the relations (\ref{qDG})
reduce to $q-$Serre relations, or for $q=1$, $\rho=\rho^*=16$ they coincide with the Dolan-Grady relations \cite{DG}.\vspace{1mm} 

The purpose of this paper is to investigate a family of TD pairs recently discovered in the context of quantum integrable systems and related algebraic structures\,\footnote{The tridiagonal algebra with relations (\ref{qDG}) is the quantum group structure behind the reflection equation associated with $U_{q^{1/2}}(\widehat{sl_2})$ \cite{qDG,TriDiag}. It also 
plays an important role as a fundamental non-abelian symmetry in some integrable systems: it generates a hierarchy of mutually commuting quantities which ensure the integrability of the system. Among the examples of integrable systems which enjoy this symmetry, one finds for instance the boundary sine-Gordon model or the XXZ open spin chain with general boundary conditions. \cite{qDG,TriDiag,qOns}.}. There are some reasons to do so \cite{Ter03}, appart from the applications to integrable systems that will be considered separately \cite{prep}. Here, we study in details their matrix representations: eigenvalue sequences, eigenvectors and block tridiagonal stucture with respect to the dual basis.  Using these data, we identify a family of coupled recurrence relations and a set of $q-$difference equations generalizing (\ref{rec0}), (\ref{qBoch}), and describe their explicit solutions in the simplest cases. Our results  are in agreement with known ones \cite{Ter01,Ter03} and provide further understanding of TD pairs. In particular, they suggest that the tridiagonal algebra (\ref{qDG}) may provide a classification scheme for orthogonal symmetric functions of one argument.\vspace{1mm}

{\bf Convention}: In this paper, ${\mathbb R}$, ${\mathbb C}$ denote the field of real and complex numbers, respectively, ${\mathbb R}^*={\mathbb R}\backslash\{0\}$, ${\mathbb C}^*={\mathbb C}\backslash\{0\}$. For convenience, the deformation parameter is sometimes denoted $q\equiv \exp(\phi)$ with $\phi\in{\mathbb C}$. According to a common notational convention, for a linear transformation ${\textsf A}$ its conjugate transpose is denoted ${\textsf A}^*$. Note that we do not use this convention. Here, the conjugate transpose of ${\textsf A}$ is denoted ${\textsf A}^\dagger$.

\section{A family of tridiagonal pairs}
Suppose that each linear transformation ${\textsf A}$, ${\textsf A}^*$ is diagonalizable on $V$ i.e $V$ is spanned by the eigenspaces of ${\textsf A}$ or ${\textsf A}^*$. 
A general object called a {\it Tridiagonal pair} is defined as follows: 
\begin{definition} \cite{Ter01}
\label{defTDpair}
Let $V$ denote
a vector space over an arbitrary field ${\mathbb K}$ with finite positive dimension. By a {\it Tridiagonal pair} (or {\it TD pair}) on $V$
we mean an ordered pair ${\textsf A}, {\textsf A}^*$, where
${\textsf A}:V\rightarrow V$ and ${\textsf A}^*:V\rightarrow V$ are linear transformations 
satisfying (i)--(iv) below.
\begin{enumerate}
\item ${\textsf A}$ and ${\textsf A}^*$ are both diagonalizable on $V$.
\item There exists an ordering $V_0, V_1,\ldots, V_d$ of the  
eigenspaces of ${\textsf A}$ such that 
\begin{equation}
{\textsf A}^* V_n \subseteq V_{n-1} + V_n+ V_{n+1} \qquad \qquad (0 \leq n \leq d),
\label{eq:tdrecall1}
\end{equation}
where $V_{-1} = 0$, $V_{d+1}= 0$.
\item There exists an ordering $V^*_0, V^*_1,\ldots, V^*_\delta$ of
the  
eigenspaces of ${\textsf A}^*$ such that 
\begin{equation}
{\textsf A} V^*_s \subseteq V^*_{s-1} + V^*_s+ V^*_{s+1} \qquad \qquad (0 \leq s \leq \delta),
\label{eq:tdrecall2}
\end{equation}
where $V^*_{-1} = 0$, $V^*_{\delta+1}= 0$.
\item There is no subspace $W$ of $V$ such  that  both ${\textsf A}W\subseteq W$,
${\textsf A}^*W\subseteq W$, other than $W=0$ and $W=V$.
\end{enumerate}
We say the above TD pair is over ${\mathbb K}$.
\end{definition}

Although a complete classification of tridiagonal pairs is not known yet, the following results have been obtained. Referring to the TD pair in the definition above,
as shown in [\cite{Ter01}, Lemma 4.2] one has $d=\delta$ which is called the {\it diameter}. Also, for $0 \leq n \leq d$, the dimensions of the eigenspaces $V_n$ and $V^*_s$ are equal and the sequence
$(dim(V_0),dim(V_1),\ldots,dim(V_d))$ - called the {\it shape vector} - is symmetric and  unimodal i.e \cite{TerIto04}
\beqa
dim(V_n) =dim(V_{d-n}) \quad \mbox{for} \quad 0 \leq n \leq d \ ,\qquad \qquad dim(V_{n-1}) \leq dim(V_{n}) \quad \mbox{for} \quad  1 \leq n \leq d/2\ .
\eeqa

Among the known examples of TD pairs, one finds a subset such that ${\textsf A},{\textsf A}^*$ have eigenspaces of dimension one i.e. with a shape vector $(1,1,\ldots,1)$.
These are called Leonard pairs, classified in \cite{TerLP01}. In particular, they satisfy (for details, see \cite{TerAW03}) the
Askey-Wilson (AW) relations (\ref{AWrel}) first introduced by Zhedanov \cite{Zhed92}. 
Other examples of TD pairs are for instance the subset corresponding to $\rho=\rho^*=0$  which reduce (\ref{qDG}) to the $q-$Serre relations of $U_{q^{1/2}}(\widehat{sl_2})$. 
In this case, the shape vector associated with the TD pair ${\textsf A}$, ${\textsf A}^*$ is such that
\beqa
dim(V_n) \leq \Biggl({{ d }\atop {n}}\Biggr)\ ,  \qquad 0 \leq n \leq d \qquad \qquad \mbox{where} \qquad \quad \Biggl({{ d }\atop {n}}\Biggr)=\frac{d!}{n!(d-n)!}
\eeqa
denotes the binomial coefficient. It is important to mention that TD pairs also arise from irreducible finite dimensional representations of the Lie algebra $sl_2$ and the Onsager algebra \cite{Ons44}. \vspace{1mm}

From now on, we focus our attention on the $U_{q^{1/2}}(\widehat{sl_2})$ algebra which plays a crucial role in the following analysis and we set ${\mathbb K}={\mathbb C}$. The quantum Kac-Moody algebra $U_{q^{1/2}}(\widehat{sl_2})$ is generated by the elements
$\{H_j,E_j,F_j\}$, $j\in \{0,1\}$. Denoting the entries of the extended Cartan matrix\,\footnote{With $i,j\in\{0,1\}$: $a_{ii}=2$,\ $a_{ij}=-2$ for $i\neq j$.} by $\{a_{ij}\}$, they satisfy the commutation
relations
\beqa [H_i,H_j]=0\ , \quad [H_i,E_j]=a_{ij}E_j\ , \quad
[H_i,F_j]=-a_{ij}F_j\ ,\quad
[E_i,F_j]=\delta_{ij}\frac{q^{H_i/2}-q^{-H_i/2}}{q^{1/2}-q^{-1/2}}\
\nonumber\eeqa
together with the $q-$Serre relations
\beqa [E_i,[E_i,[E_i,E_j]_{q}]_{q^{-1}}]=0\ ,\quad \mbox{and}\quad
[F_i,[F_i,[F_i,F_j]_{q}]_{q^{-1}}]=0\ . \label{defUq}\eeqa
The sum ${\it K}=H_0+H_1$ is the central element of the algebra. The
Hopf algebra structure is ensured by the existence of a
comultiplication $\Delta: U_{q^{1/2}}(\widehat{sl_2})\rightarrow U_{q^{1/2}}(\widehat{sl_2})\otimes U_{q^{1/2}}(\widehat{sl_2})$ 
and a counit ${\cal E}: U_{q^{1/2}}(\widehat{sl_2})\rightarrow {\mathbb C}$ with
\beqa \Delta(E_i)&=&E_i\otimes q^{H_i/4} +
q^{-H_i/4}\otimes E_i\ ,\nonumber \\
 \Delta(F_i)&=&F_i\otimes q^{H_i/4} + q^{-H_i/4}\otimes F_i\ ,\nonumber\\
 \Delta(H_i)&=&H_i\otimes I\!\!I + I\!\!I \otimes H_i\ \label{coprod}
\eeqa
and\vspace{-0.3cm}
\beqa {\cal E}(E_i)={\cal E}(F_i)={\cal
E}(H_i)=0\ ,\qquad {\cal E}({I\!\!I})=1\
.\label{counit}\nonumber\eeqa
More generally, one defines the $N-$coproduct $\Delta^{(N)}: \
U_{q^{1/2}}(\widehat{sl_2}) \longrightarrow
U_{q^{1/2}}(\widehat{sl_2}) \otimes \cdot\cdot\cdot \otimes
U_{q^{1/2}}(\widehat{sl_2})$ as
\beqa \Delta^{(N)}\equiv (id\times \cdot\cdot\cdot \times id
\times \Delta)\circ \Delta^{(N-1)}\ \label{coprodN}\eeqa
for $N\geq 3$ with $\Delta^{(2)}\equiv \Delta$,
$\Delta^{(1)}\equiv id$. The opposite $N-$coproduct
$\Delta'^{(N)}$ is similarly defined with $\Delta'\equiv \sigma
\circ\Delta$   where the permutation map $\sigma(x\otimes y
)=y\otimes x$ for all $x,y\in U_{q^{1/2}}(\widehat{sl_2})$ is
used. 
%
%
%
\vspace{1mm}

In order to construct explicitely a family of TD pairs, we use the link recently exhibited \cite{qDG,TriDiag} between the tridiagonal algebra 
$\mathbb{T}$ and a class of quadratic algebra, namely the reflection equation introduced in \cite{Cher,Skly}. Based on \cite{qDG,TriDiag},
we restrict our attention to the quantum Kac-Moody algebra $U_{q^{1/2}}(\widehat{sl_2})$
with zero center i.e. ${\it K}\equiv 0$. In this special case, this algebra with defining relations (\ref{defUq}) is called the quantum loop algebra of $sl_2$ denoted $U_{q^{1/2}}({\cal L }(sl_2))$ below. We have the following result (see also \cite{qDG,TriDiag}):
\begin{proposition}
\label{p1}
Let $\{k_+,k_-,\epsilon_\pm\}$ denote nonzero scalars in ${\mathbb C}$. There is an algebra homomorphism ${\mathbb T}\mapsto U_{q^{1/2}}({\cal L }(sl_2))$ such that
\beqa
{\textsf A}&\mapsto& k_+E_1q^{H_1/4} + k_-F_1q^{H_1/4}+\epsilon_+ q^{H_1/2}\nonumber\\
{\textsf A}^*&\mapsto& k_-E_0q^{H_0/4} + k_+F_0q^{H_0/4}+\epsilon_- q^{H_0/2}\label{defhA}
\eeqa
with
\beqa
\rho=\rho^*=(q^{1/2}+q^{-1/2})^2k_+k_-\ .\label{rho}
\eeqa
\end{proposition}

{\it Proof:} Replacing (\ref{defhA}) in (\ref{qDG}), we use the defining relations and the $q-$Serre relations (\ref{defUq}) to 
simplify the l.h.s. of the tridiagonal relations. Straightforward calculations show that the l.h.s. of (\ref{qDG})
reduces to the r.h.s provided $\rho=\rho^*$ is fixed in terms of the deformation parameter $q$, with the relation given above.\finproof
\begin{remark}
\label{}
This realization is different from the one
proposed by Ito and Terwilliger in \cite{TerIto04} in which case
$\rho=0$, i.e. (\ref{qDG}) reduce to $q-$Serre relations.
\end{remark}
\begin{remark}
\label{}
Using the homorphism defined above, it follows \cite{TriDiag}
that the corresponding tridiagonal relations (\ref{qDG}) are invariant under the action of the coproduct (\ref{coprod}) as well as its generalization (\ref{coprodN}).
\end{remark}

To obtain finite dimensional irreducible representations of the standard generators (\ref{defhA}), one introduces the evaluation homomorphism $\pi_v: U_{q^{1/2}}({\cal L }(sl_2))\mapsto U_{q^{1/2}}(sl_2))$ in the principal gradation:   
\beqa &&\pi_v[E_1]= vS_+\ , \qquad \ \ \ \pi_v[E_0]= vS_-\ , \nonumber\\
&&\pi_v[F_1]=
v^{-1}S_-\ ,\qquad \pi_v[F_0]= v^{-1}S_+\ ,\nonumber\\
\ &&\pi_v[q^{H_1/2}]= q^{s_3}\ ,\qquad \ \pi_v[q^{H_0/2}]=
q^{-s_3}\ ,\nonumber\eeqa
where the generators of $U_{q^{1/2}}(sl_2)$ satisfy
\beqa
[s_3,S_\pm]=\pm S_\pm \qquad \mbox{and} \qquad
[S_+,S_-]=(q^{s_3}-q^{-s_3})/(q^{1/2}-q^{-1/2})\ .\nonumber
\eeqa
As shown in \cite{Chari91}, every finite dimensional irreducible representation of $U_{q^{1/2}}({\cal L }(sl_2))$ is a tensor product of evaluation representations. Using the $N-$coproduct (\ref{coprodN}), it follows \cite{qDG,TriDiag}:
\begin{lemma}
\label{}
Let $k_+=(k_-)^\dagger,\epsilon_\pm$ nonzero scalars in ${\mathbb C}$. For any $v_i\in {\mathbb C}^*$ with integers $i=1,...,N$, let the pair of elements ${\textsf W}_{0}^{(N)},{\textsf W}_{1}^{(N)}$ defined by
\beqa
{\textsf W}_{0}^{(N)}&=&
\left(k_+v_Nq^{1/4}S_+q^{s_3/2}
+k_-v_N^{-1}q^{-1/4}S_-q^{s_3/2}\right)\otimes I\!\!I^{(N-1)}\ +\ q^{s_3}\otimes {\textsf W}_{0}^{(N-1)}
\ ,\nonumber\\
{\textsf W}_{1}^{(N)}&=&
\left(k_+v^{-1}_Nq^{-1/4}S_+q^{-s_3/2} + k_-v_Nq^{1/4}S_-q^{-s_3/2}\right)\otimes I\!\!I^{(N-1)}\ +\ q^{-s_3}\otimes {\textsf W}_{1}^{(N-1)}
\ \label{TDpairgen}
\eeqa
with \ ${{\textsf W}}_{0}^{(0)}\equiv \epsilon_+\ , {{\textsf W}}_{1}^{(0)}\equiv \epsilon_-$. 
Let $V$ denote a finite dimensional $N-$tensor product representation of $\otimes_1^N U_{q^{1/2}}({\cal L }(sl_2))$. Assume $V$ is an irreducible $({\textsf W}_{0}^{(N)},{\textsf W}_{1}^{(N)})-$module and each of ${\textsf W}_{0}^{(N)},{\textsf W}_{1}^{(N)}$ is diagonalizable on $V$. Then the pair ${\textsf W}_{0}^{(N)},{\textsf W}_{1}^{(N)}$ acts on $V$ as a tridiagonal pair. 
\end{lemma}

{\it Proof}: Conditions 1 and 4 of Definition \ref{defTDpair} hold by assumption. From Proposition \ref{p1}, in the evaluation representation of  $U_{q^{1/2}}({\cal L }(sl_2))$ the elements ${\textsf W}_{0}^{(1)},{\textsf W}_{1}^{(1)}$ satisfy the tridiagonal relations (\ref{defhA}). Using the $N-$coproduct (\ref{coprodN}) and the result of \cite{Chari91}, it follows that ${\textsf W}_{0}^{(N)},{\textsf W}_{1}^{(N)}$ satisfy the tridiagonal relations (\ref{defhA}) with (\ref{rho}) for general values of $N$. Then, according to [\cite{Ter03}, Theorem 10] the assertion follows.\finproof

\section{Matrix representation and block tridiagonal structure}
For general values of $q$, irreducible  $(2j+1)-$finite dimensional representations of  $U_{q^{1/2}}(sl_2)$ are well-known. In the following, for simplicity we will focus on the case $j=1/2$ and introduce the canonical basis\,\footnote{The scalar product $\langle . , .\rangle$ is defined by $\langle f^j_\pm, f^j_\pm\rangle\equiv 1$, $\langle f^j_\pm, f^j_\mp\rangle\equiv 0$ for any integer $j$.} $\{f^i_+,f^i_-\}$ for the $i-th$ two-dimensional representation in the $N-$tensor product representation of $\otimes_1^N U_{q^{1/2}}(sl_2)$. Defining the endomorphism $\xi: U_{q^{1/2}}(sl_2)\mapsto End({\mathbb C}^2)$, in this basis one has in terms of the Pauli matrices $\sigma_\pm,\sigma_3$:
\beqa
S_+\mapsto\sigma_+=\left(
\begin{array}{cc}
 0    & 1 \\
 0 & 0 
\end{array} \right) \ ,\qquad
S_-\mapsto\sigma_- =\left(
\begin{array}{cc}
 0    & 0 \\
 1 & 0 
\end{array} \right) \ ,\qquad
2s_3\mapsto\sigma_3 =\left(
\begin{array}{cc}
 1    & 0 \\
 0 & -1 
\end{array} \right) \ .\label{Pauli}
\eeqa
\begin{definition}
\label{}
Define $k_+=(k_-)^\dagger=-(q^{1/2}-q^{-1/2})e^{i\theta}/2$ with $\theta\in{\mathbb R}$, $q=e^\phi$ with $\phi$ purely imaginary, and $\{\alpha,\alpha^*\}\in {\mathbb C}^*$. Let \ $V=({\mathbb C}^{2})^{\otimes N}$. Define ${\cal W}^{(0)}_0\equiv \epsilon_+=\cosh\alpha$, ${\cal W}^{(0)}_1\equiv \epsilon_-=\cosh\alpha^*$. We define the family of matrices of size $2^N\times 2^N$:
\beqa
{\cal W}^{(N)}_0&=& (k_+\sigma_+ + k_-\sigma_-)\otimes I\!\!I^{(N-1)} + q^{\sigma_3/2}\otimes {\cal W}_0^{(N-1)}\ ,\nonumber \\
{\cal W}^{(N)}_1&=& (k_+\sigma_+ + k_-\sigma_-)\otimes I\!\!I^{(N-1)} + q^{-\sigma_3/2}\otimes {\cal W}_1^{(N-1)}\ .\label{op}
\eeqa
\end{definition}
Below, we study in details the structure of these matrices for general values of $N$ in their eigenbasis and dual. Although it corresponds to the spin-$j=1/2$ case, the structure of the pair ${\cal W}^{(N)}_0,{\cal W}^{(N)}_1$ can be studied along the same line for higher representations leading to analogous results.

\subsection{Example $N=1$ and the Askey-Wilson algebra}
As pointed out in \cite{qDG}, for $N=1$ the elements ${\textsf W}^{(1)}_0,{\textsf W}^{(1)}_1$ satisfy the Askey-Wilson algebra (\ref{AWrel}) introduced in \cite{Zhed92}. In particular, their matrix representations of size $2j+1\times 2j+1$ provide an example of Leonard pairs \cite{TerAW03} with shape vector $(1,1,...,1)$.   
For the simplest case $j=1/2$, ${\cal W}^{(1)}_0,{\cal W}^{(1)}_1$ are $2\times 2$ matrices which entries in the canonical basis follow from (\ref{op}) with (\ref{Pauli}). Let $\psi^{(1)}_{n}$, $n=0,1$ denote the two eigenvectors  of ${\cal W}^{(1)}_0$ in $V$ with (multiplicity free) eigenvalues $\lambda^{(1)}_n$.  One finds $\lambda^{(1)}_n=\cosh(\alpha+(1-2n)\phi/2)$, $n=0,1$ with
\beqa
\psi^{(1)}_{0}= e^{\alpha+i\theta}f^1_+ + f^1_-\ ,\qquad \psi^{(1)}_{1}= e^{-\alpha+i\theta}f^1_+ + f^1_-\ .\label{eigenvect1}
\eeqa
Assuming $\alpha$ generic, they are independent of each other. In this new basis, the element ${\cal W}^{(1)}_1$ can be written as
\beqa
{\cal W}^{(1)}_1=\left(
\begin{array}{cc}
   \acal^{(1)}_0  & \ccal^{(1)}_1 \\
 \bcal^{(1)}_0 & \acal^{(1)}_1 
\end{array} \right) \ ,\label{init}
\eeqa
where
\beqa
\acal^{(1)}_0&=&\frac{\cosh\alpha^*\sinh(\alpha-\phi/2)-\sinh(\phi/2)}{\sinh\alpha},\qquad
\acal^{(1)}_1=\frac{\cosh\alpha^*\sinh(\alpha+\phi/2)+\sinh(\phi/2)}{\sinh\alpha}, \nonumber\\
\bcal^{(1)}_0&=&\frac{e^\alpha(\cosh\alpha+\cosh\alpha^*)\sinh(\phi/2)}{\sinh\alpha}, \qquad \quad
\ccal^{(1)}_1=-\frac{e^{-\alpha}(\cosh\alpha+\cosh\alpha^*)\sinh(\phi/2)}{\sinh\alpha}\ . \nonumber
\eeqa

By analogy, for generic $\alpha^*\in {\mathbb C}$ the eigenvectors of the element ${\cal W}^{(1)}_1$ - denoted $\varphi^{(1)}_{s}$ with $s=0,1$ - form a complete basis of $V$. They are easily obtained substituting 
\beqa
n\rightarrow s\ , \quad\alpha\leftrightarrow -\alpha^*\ \ , \quad \phi\rightarrow -\phi\ \ , \quad \theta\rightarrow \theta+\pi\label{subs}
\eeqa
in (\ref{eigenvect1}). In this dual basis, the matrix ${\cal W}^{(1)}_1$ is diagonal with eigenvalues $\lambda^{(1)}_s=\cosh(\alpha^*+(1-2s)\phi/2)$, $s=0,1$ whereas
the entries $\{\acalt^{(1)}_s, \bcalt^{(1)}_s, \ccalt^{(1)}_s\}$ of ${\cal W}^{(1)}_0$ follow from $\{\acal^{(1)}_n, \bcal^{(1)}_n, \ccal^{(1)}_n\}$ using (\ref{subs}).  

\subsection{Example $N=2$}
Contrary to the case $N=1$, the elements ${\textsf W}^{(2)}_0,{\textsf W}^{(2)}_1$ do not satisfy the defining relations (\ref{AWrel}) of the Askey-Wilson algebra \cite{TriDiag}. Instead, they satisfy a pair of fifth order relations that can be found in \cite{qOns}. Also, the eigenvalues of the $4\times 4$ dimensional matrices ${\cal W}^{(2)}_0,{\cal W}^{(2)}_1$ are not mutually distinct. More precisely, the shape vector associated with these eigenvalues $\lambda^{(2)}_n=\cosh(\alpha+(2-2n)\phi/2)$ for $n=0,1,2$ takes the form  $(dim(V_0),dim(V_1),dim(V_2))=(1,2,1)$. For generic values of the parameters $\alpha,\phi,\theta$ the corresponding eigenvectors denoted $\psi^{(2)}_{n[i]}$ form a complete basis.  In the canonical basis $(f_\pm^2\otimes f_\pm^1)$ of $U_{q^{1/2}}(sl_2)\otimes U_{q^{1/2}}(sl_2)$, they take the form
\beqa
\psi^{(2)}_{0[1]}&=& (e^{\alpha+\phi/2+i\theta}f^2_+ + f^2_-)\otimes (e^{\alpha+i\theta}f^1_+ + f^1_-)\ ,\nonumber\\
\psi^{(2)}_{1[1]}&=& (e^{\alpha-\phi/2+i\theta}f^2_+ + f^2_-)\otimes (e^{-\alpha+i\theta}f^1_+ + f^1_-)\ ,\nonumber\\
\psi^{(2)}_{1[2]}&=& (e^{-\alpha-\phi/2+i\theta}f^2_+ + f^2_-)\otimes (e^{\alpha+i\theta}f^1_+ + f^1_-)\ ,\nonumber\\
\psi^{(2)}_{2[1]}&=& (e^{-\alpha+\phi/2+i\theta}f^2_+ + f^2_-)\otimes (e^{-\alpha+i\theta}f^1_+ + f^1_-)\ ,\label{eigenvect2}
\eeqa
where the index $i=1,2$ which characterizes the multiplicity of the eigenvalues has been introduced. With respect to the basis $\psi^{(2)}_{0[1]},\psi^{(2)}_{1[1]},\psi^{(2)}_{1[2]},\psi^{(2)}_{2[1]}$ the matrices representing 
${\cal W}^{(2)}_0,{\cal W}^{(2)}_1$ read, respectively,
\beqa
{\cal W}^{(2)}_0=\mbox{diag}(\lambda^{(2)}_0,\lambda^{(2)}_1,\lambda^{(2)}_1,\lambda^{(2)}_2)\ ,\qquad {\cal W}^{(2)}_1=\left(
\begin{array}{cccc}
   \acal^{(2)}_{0[11]}  & \ccal^{(2)}_{1[11]} & \ccal^{(2)}_{1[12]}  & 0\\
   \bcal^{(2)}_{0[11]} & \acal^{(2)}_{1[11]} & \acal^{(2)}_{1[12]} & \ccal^{(2)}_{2[11]} \\
   \bcal^{(2)}_{0[21]} & \acal^{(2)}_{1[21]} & \acal^{(2)}_{1[22]} & \ccal^{(2)}_{2[21]} \\
   0 & \bcal^{(2)}_{1[11]} & \bcal^{(2)}_{1[12]} & \acal^{(2)}_{2[11]} \label{mat2}
\end{array} \right) \ ,
\eeqa
where the entries are
\beqa
\acal^{(2)}_{0[11]}&=&\frac{\acal^{(1)}_0\sinh\alpha-\sinh(\phi/2)}{\sinh(\alpha+\phi/2)}, \nonumber \\
\acal^{(2)}_{1[11]}&=&\frac{\acal^{(1)}_1\sinh(\alpha-\phi)-\sinh(\phi/2)}{\sinh(\alpha-\phi/2)}, \ \ \qquad \acal^{(2)}_{1[12]}=-\frac{e^{-\alpha}\sinh\phi}{\sinh(\alpha-\phi/2)}\bcal^{(1)}_0, \nonumber \\
\acal^{(2)}_{1[21]}&=&\frac{e^{\alpha}\sinh\phi}{\sinh(\alpha+\phi/2)}\ccal^{(1)}_1, \qquad \qquad \qquad \quad \acal^{(2)}_{1[22]}=\frac{\acal^{(1)}_0\sinh(\alpha+\phi)+\sinh(\phi/2)}{\sinh(\alpha+\phi/2)}, \nonumber \\
\acal^{(2)}_{2[11]}&=&\frac{\acal^{(1)}_1\sinh\alpha+\sinh(\phi/2)}{\sinh(\alpha-\phi/2)}, \nonumber \\
\bcal^{(2)}_{0[11]}&=&e^{\phi/2}\bcal^{(1)}_0, \qquad \qquad 
\bcal^{(2)}_{0[21]}=\frac{e^{\alpha+\phi/2}\sinh(\phi/2)}{\sinh(\alpha+\phi/2)}\big(\acal^{(1)}_0+\cosh(\alpha+\phi/2)\big), \nonumber\\
\bcal^{(2)}_{1[11]}&=&\frac{e^{\alpha-\phi/2}\sinh(\phi/2)}{\sinh(\alpha-\phi/2)}\big(\acal^{(1)}_1+\cosh(\alpha-\phi/2)\big), \qquad 
\bcal^{(2)}_{1[12]}=e^{-\phi/2}\frac{\sinh(\alpha+\phi/2)}{\sinh(\alpha-\phi/2)}\bcal^{(1)}_0, \nonumber\\
\ccal^{(2)}_{2[11]}&=&-\frac{e^{-\alpha+\phi/2}\sinh(\phi/2)}{\sinh(\alpha-\phi/2)}\big(\acal^{(1)}_1+\cosh(\alpha-\phi/2)\big), \qquad
\ccal^{(2)}_{2[21]}=e^{\phi/2}\ccal^{(1)}_1, \nonumber\\
\ccal^{(2)}_{1[11]}&=&e^{-\phi/2}\frac{\sinh(\alpha-\phi/2)}{\sinh(\alpha+\phi/2)}\ccal^{(1)}_1, \qquad
\ccal^{(2)}_{1[12]}=-\frac{e^{-\alpha-\phi/2}\sinh(\phi/2)}{\sinh(\alpha+\phi/2)}\big(\acal^{(1)}_0+\cosh(\alpha+\phi/2)\big)\ .\nonumber
\eeqa

The set of eigenvectors $\varphi^{(2)}_{s[k]}$ that diagonalize ${\cal W}^{(1)}_1$  with $s=0,2$ $(k=1)$ and $s=1$ $(k=1,2)$  are obtained using (\ref{subs}) in (\ref{eigenvect2}), and their eigenvalues are $\lambda^{(2)}_s=\cosh(\alpha^*+(2-2s)\phi/2)$, $s=0,1,2$.  In this dual basis the matrix ${\cal W}^{(2)}_0$ takes a block tridiagonal form similar to ${\cal W}^{(2)}_1$ in (\ref{mat2}), where the corresponding entries $\{\acalt^{(2)}_s, \bcalt^{(2)}_s, \ccalt^{(2)}_s\}$ follow from $\{\acal^{(2)}_n, \bcal^{(2)}_n, \ccal^{(2)}_n\}$ using (\ref{subs}).
Assuming that there are no special relations between the parameters $\alpha,\phi,\theta$ or $\alpha^*,\phi,\theta$, it follows from above results that the matrices ${\cal W}^{(2)}_0$, ${\cal W}^{(2)}_1$ satisfy the requirements of Definition \ref{defTDpair}. The block tridiagonal structure, which didn't appear for $N=1$ is now clear from (\ref{mat2}). In the next Section, we show that this property generalizes for higher values of $N$.   

\subsection{Generalization}
To describe the family of matrices ${\cal W}^{(N)}_0,{\cal W}^{(N)}_1$ in details, first we need to construct explicitely a complete basis of eigenvectors $\psi^{(N)}_{n[i]}$ that diagonalize ${\cal W}^{(N)}_0$ for any $N$, and identify the corresponding eigenvalues. 
\begin{definition}
\label{}
Let $V$ denote the finite dimensional representation (of dimension $2^N$) on which ${\cal W}_{0}^{(N)}$ acts. Suppose ${\cal W}_{0}^{(N)}$ is diagonalizable on $V$, and let $V_0,V_1,V_2,...$ denote a standard ordering of the eigenspaces of ${\cal W}_{0}^{(N)}$.
For $n=0,1,2,...$, let $\lambda^{(N)}_{n}$ denote the eigenvalue sequence of ${\cal W}_{0}^{(N)}$.
\end{definition}
The structure of the eigenvectors and eigenvalues is as follows:
\begin{proposition}
\label{diagW0}
Define $k_+=(k_-)^\dagger=-(q^{1/2}-q^{-1/2})e^{i\theta}/2$. Let $\{\alpha,\alpha^*\}\in{\mathbb C}^*$, $\theta\in{\mathbb R}$ be generic scalars and $\phi$ purely imaginary. Introduce the set $\epsilon^{[i]}_j=\pm 1$, $j=1,...,N$ and $i\in\{1,...,dim(V_n)\}$. The matrix ${\cal W}_{0}^{(N)}$ defined in (\ref{op}) is diagonalized by the eigenvectors
\beqa
\psi^{(N)}_{n[i]}=\bigotimes_{l=1}^N \left( e^{\epsilon^{[i]}_l\alpha+\epsilon^{[i]}_l\sum_{k=1}^{l-1}\epsilon^{[i]}_k\phi/2+i\theta}f^l_+ + f^l_- \right)\ .\label{eigenvectN}
\eeqa
The corresponding eigenvalues, ordered by the integer $n=0,1,...,N$, are given by
\beqa
\lambda^{(N)}_{n}=\cosh(\alpha+(N-2n)\phi/2) \qquad \mbox{where} \qquad n=(N-\sum_{k=1}^{N}\epsilon^{[i]}_k)/2\ .\label{eigenvalN}
\eeqa
\end{proposition}

{\it Proof}: According to previous results, the assertion is true for $N=1$, $N=2$. To show it for all $N$, we proceed by recursion. Suppose
the assertion holds for $N$ fixed i.e   ${\cal W}_{0}^{(N)}\psi^{(N)}_{n[i]}=\lambda^{(N)}_{n}\psi^{(N)}_{n[i]}$ for a given set $\epsilon^{[i]}_j=\pm 1$. Replacing $N\rightarrow N+1$ in the definition (\ref{op}) and using 
\beqa
\sigma_\pm f^{N+1}_\mp=f^{N+1}_\pm\ ,\quad \sigma_\pm f^{N+1}_\pm=0\ ,\quad \sigma_3 f^{N+1}_\pm=\pm f^{N+1}_\pm\ ,\nonumber
\eeqa
we obtain
\beqa
&&{\cal W}_{0}^{(N+1)}\Big( e^{\epsilon^{[i]}_{N+1}\alpha+\epsilon^{[i]}_{N+1}\sum_{k=1}^{N}\epsilon^{[i]}_k\phi/2+i\theta}f^{N+1}_+ + f^{N+1}_- \Big)\otimes \psi^{(N)}_{n[i]}\ \nonumber\\
&&\qquad \qquad \qquad = \Big( (k_++q^{1/2}\lambda^{(N)}_{n}e^{\epsilon^{[i]}_{N+1}\alpha+\epsilon^{[i]}_{N+1}\sum_{k=1}^{N}\epsilon^{[i]}_k\phi/2+i\theta})f^{N+1}_+ \nonumber\\
&&\qquad \qquad \qquad \qquad + (k_-e^{\epsilon^{[i]}_{N+1}\alpha+\epsilon^{[i]}_{N+1}\sum_{k=1}^{N}\epsilon^{[i]}_k\phi/2+i\theta})+q^{-1/2}\lambda^{(N)}_{n})f^{N+1}_- \Big)\otimes \psi^{(N)}_{n[i]}\ .\nonumber
\eeqa 
From the definition of $k_\pm$ together with (\ref{eigenvalN}) and the explicit expression of $n$, the r.h.s. of the equation above reduces to
\beqa
\cosh\big(\alpha+\big(\sum_{k=1}^{N}\epsilon^{[i]}_k + \epsilon^{[i]}_{N+1}\big)\phi/2\big)\ \Big( e^{\epsilon^{[i]}_{N+1}\alpha+\epsilon^{[i]}_{N+1}\sum_{k=1}^{N}\epsilon^{[i]}_k\phi/2+i\theta}f^{N+1}_+ + f^{N+1}_- \Big)\otimes \psi^{(N)}_{n[i]}\ ,
\eeqa
provided  $\epsilon^{[i]}_{N+1}=\pm 1$. Identifying $\sum_{k=1}^{N+1}\epsilon^{[i]}_k=N+1-2n$, the eigenvalues take the form proposed in (\ref{eigenvalN}) replacing $N$ by $N+1$. As a consequence, the eigenvectors and eigenvalues structure for $N$ also holds for $N+1$. The assertion being true for $N=1,2$, it holds for all $N$.\finproof
\begin{corollary}
\label{nombreval}
Let $\{\alpha,\alpha^*\}\in{\mathbb C}^*$, $\theta\in{\mathbb R}$ be generic scalars and $\phi$ purely imaginary.
The matrix ${\cal W}_{0}^{(N)}$ has exactly $N+1$ mutually distinct eigenvalues $\lambda^{(N)}_{n}$. The dimension of an eigenspace $V_n$ is:
\beqa
dim(V_n) = \Biggl({{ N }\atop {n}}\Biggr)\label{dimVn}\ .
\eeqa
\end{corollary}

{\it Proof}: First, according to the definition of $n$ in (\ref{eigenvalN}) and $\epsilon^{[i]}_k=\pm 1$, one has $-N\leq \sum_{k=1}^{N}\epsilon^{[i]}_k \leq N$. It yields $0 \leq n\leq N$, which shows the first part of the assertion. To show (\ref{dimVn}), we proceed by recursion. For $N=1$, $N=2$ the assertion is true. Let $N$ be fixed and suppose (\ref{dimVn}). Notice that $dim(V_n)$ is the total number of configurations with index ${[i]}$ characterized by the sequence $(\epsilon^{[i]}_1=\pm 1,...,\epsilon^{[i]}_N=\pm 1)$. Replacing $N$ by $N+1$ in (\ref{eigenvectN}) and setting $n=(N+1-\sum_{k=1}^{N+1}\epsilon^{[i]}_k)/2$ there are two families of configurations associated with $n$:
\beqa
&&\big(\underbrace{(\epsilon^{[i]}_1=\pm 1,...,\epsilon^{[i]}_j=\pm 1,\epsilon^{[i]}_{j+1}=\mp 1,...,\epsilon^{[i+1]}_{N}=\pm 1)}_{\big({{ N }\atop {n-1}}\big) \quad \mbox{configurations} },\epsilon^{[i]}_{N+1}= +1\big)\nonumber\\
\mbox{and}\quad && \big(\underbrace{(\epsilon^{[i]}_1=\pm 1,...,\epsilon^{[i]}_k=\pm 1,\epsilon^{[i]}_{k+1}=\pm 1,...,\epsilon^{[i+1]}_{N}=\pm 1)}_{\big({{ N }\atop {n}}\big) \quad \mbox{configurations} },\epsilon^{[i]}_{N+1}= -1\big)\ .\label{config}
\eeqa  
It follows that $V_n$ for $N+1$ is such that $dim(V_n) =\big({{ N }\atop {n-1}}\big)+\big({{ N }\atop {n}}\big)=\big({{ N+1 }\atop {n}}\big)$ in agreement with (\ref{dimVn}). Then, (\ref{dimVn}) holds for all values of $N$.\finproof   \vspace{0.3cm}

For {\it generic} scalars $\{\alpha,\alpha^*,\phi,\theta\}$, the eigenvectors (\ref{eigenvectN}) are linearly independent. In addition, $\sum_{n=0}^N dim(V_n) = 2^N=dim(V)$  so they provide a complete basis for $V$.  Let us now show that ${\cal W}^{(N)}_1$ exhibits a block tridiagonal structure in the basis $\psi^{(N)}_{n[i]}$, $i=1,2,...,\big({{ N }\atop {n}}\big)$. 
\begin{proposition}
\label{tridiagW1} 
Define $k_+=(k_-)^\dagger=-(q^{1/2}-q^{-1/2})e^{i\theta}/2$. Let $\{\alpha,\alpha^*\}\in{\mathbb C}^*$, $\theta\in{\mathbb R}$ be generic scalars and $\phi$ purely imaginary. Let $\{\psi^{(N)}_{n[j]}\}$, $j=1,2,...,\big({{ N }\atop {n}}\big)$ as defined by (\ref{eigenvectN}) the basis such that
\beqa
{\cal W}^{(N)}_0\psi^{(N)}_{n[j]}&=&\lambda^{(N)}_n\psi^{(N)}_{n[j]} \qquad \mbox{with} \qquad \lambda^{(N)}_n=\cosh(\alpha+(N-2n)\phi/2)\label{specvalN}
\eeqa
for $n=0,1,...,N$. Then, the matrix ${\cal W}_{1}^{(N)}$ acts on $V_n$ as: 
\beqa
{\cal W}^{(N)}_1\psi^{(N)}_{n[j]}&=& \sum_{i=1}^{\binombN}\bcal^{(N)}_{n[ij]}\psi^{(N)}_{n+1[i]} + \sum_{i=1}^{\binomaN}\acal^{(N)}_{n[ij]}\psi^{(N)}_{n[i]} + \sum_{i=1}^{\binomcN}\ccal^{(N)}_{n[ij]}\psi^{(N)}_{n-1[i]} \ ,\label{recpsi}
\eeqa
where the coefficients $\acal^{(N)}_{n[ij]},\bcal^{(N)}_{n[ij]},\ccal^{(N)}_{n[ij]}$ are determined recursively from (\ref{init}).
\end{proposition}

{\it Proof}: For $N=1,2$, the assertion holds, with the entries given by (\ref{init}) and (\ref{mat2}), respectively. To show the assertion for all $N$, we proceed by recursion. Let $N$ be fixed and suppose (\ref{recpsi}) is true. As noticed above, given $n$ there are two families of possible configurations, given by (\ref{config}). Replacing $N$ by $N+1$ in (\ref{eigenvectN}) the corresponding family of eigenvectors reads:
\beqa
\psi^{(N+1)}_{n[j]}&=&\Big( e^{\alpha+(N-2n)\phi/2+i\theta}f^{N+1}_+ + f^{N+1}_- \Big)\otimes \psi^{(N)}_{n[j]}\ \quad \quad\ \ \quad \quad\quad \mbox{for}\quad j\in \{1,...,\Big({{N}\atop {n}}\Big)\}\ ,\nonumber\\
\psi^{(N+1)}_{n[j]}&=&\Big(e^{-\alpha-(N-2n+2)\phi/2+i\theta}f^{N+1}_+ + f^{N+1}_- \Big)\otimes \psi^{(N)}_{n-1[j-\binomaN]}\ \quad \mbox{for}\quad j\in \{\Big({{N}\atop {n}}\Big)+1,...,\Big({{N+1}\atop {n}}\Big)\}\ .\nonumber
\eeqa
Having these equations, we can now consider the action of ${\cal W}^{(N+1)}_1$ on the eigenvectors, using (\ref{op}). Let us first consider the domain $j\in \{1,...,\big({{ N }\atop {n}}\big)\}$:
\beqa
{\cal W}_{1}^{(N+1)} \psi^{(N+1)}_{n[j]}\! \!&=& \!\!\Big( \big(k_+ + \acal^{(N)}_{n[jj]} q^{-1/2}e^{\alpha+(N-2n)\phi/2+i\theta}\big)f^{N+1}_+ 
+ \big(k_-e^{\alpha+(N-2n)\phi/2+i\theta}+q^{1/2}\acal^{(N)}_{n[jj]}\big)f^{N+1}_- \Big)\otimes \psi^{(N)}_{n[j]}\nonumber\\
&&+\sum_{i=1,i\neq j}^{\binomN} \Big(\big(\acal^{(N)}_{n[ij]}q^{-1/2}e^{\alpha+(N-2n)\phi/2+i\theta}\big)f^{N+1}_+ +  \big(\acal^{(N)}_{n[ij]}q^{1/2}\big)f^{N+1}_-\Big)\otimes \psi^{(N)}_{n[i]}\nonumber\\
&&+ \sum_{i=1}^{\binomNpun}\Big(\big(\bcal^{(N)}_{n[ij]}q^{-1/2}e^{\alpha+(N-2n)\phi/2+i\theta}\big)f^{N+1}_+ +  \big(\bcal^{(N)}_{n[ij]}q^{1/2}\big)f^{N+1}_-\Big)\otimes \psi^{(N)}_{n+1[i]}\nonumber\\
&&+\sum_{i=1}^{\binomNnmun}\Big(\big(\ccal^{(N)}_{n[ij]}q^{-1/2}e^{\alpha+(N-2n)\phi/2+i\theta}\big)f^{N+1}_+ +  \big(\ccal^{(N)}_{n[ij]}q^{1/2}\big)f^{N+1}_-\Big)\otimes \psi^{(N)}_{n-1[i]}\ .
\label{eq1}
\eeqa
On the other hand, ${\cal W}^{(N+1)}_1$ is obtained replacing $N$ by $N+1$ in (\ref{recpsi}). It must coincide with
\beqa
{\cal W}^{(N+1)}_1\psi^{(N+1)}_{n[j]}&=& \sum_{i=1}^{\left(^{N+1}_{n+1}\right)}\bcal^{(N+1)}_{n[ij]}\psi^{(N)}_{n+1[i]} + \sum_{i=1}^{\left(^{N+1}_{n}\right)}\acal^{(N+1)}_{n[ij]}\psi^{(N+1)}_{n[i]} + \sum_{i=1}^{\left(^{N+1}_{n-1}\right)}\ccal^{(N+1)}_{n[ij]}\psi^{(N+1)}_{n-1[i]} \ .
\label{eq2}
\eeqa
Taking into account the two families of eigenvectors and changing appropriately the indices, (\ref{eq2}) becomes
\beqa
{\cal W}_{1}^{(N+1)} \psi^{(N+1)}_{n[j]}\! \!&=& \!\!\sum_{i=1}^{\binomN} \Big( \big(\acal^{(N+1)}_{n[ij]}e^{\alpha+(N-2n)\phi/2+i\theta}+\bcal^{(N+1)}_{n[i+\binomNpun\ j]}e^{-\alpha-(N-2n)\phi/2+i\theta}\big)f^{N+1}_+\nonumber\\ 
&&\qquad \qquad \qquad\qquad\qquad\qquad\qquad+ \big(\acal^{(N+1)}_{n[ij]} + \bcal^{(N+1)}_{n[i+\binomNpun\ j]}\big)f^{N+1}_- \Big)\otimes \psi^{(N)}_{n[i]}\nonumber\\
&&+\sum_{i=1}^{\binomNpun} \Big( \big(\bcal^{(N+1)}_{n[ij]}e^{\alpha+(N-2n-2)\phi/2+i\theta}\big)f^{N+1}_+ + \big(\bcal^{(N+1)}_{n[ij]}\big)f^{N+1}_- \Big) \otimes \psi^{(N)}_{n+1[i]}\nonumber\\
&&+ \sum_{i=1}^{\binomNnmun}\Big( \big(\acal^{(N+1)}_{n[i+\binomaN\ j]}e^{-\alpha-(N-2n+2)\phi/2+i\theta}+\ccal^{(N+1)}_{n[ij]}e^{\alpha+(N-2n+2)\phi/2+i\theta}\big)f^{N+1}_+\nonumber\\ 
&&\qquad \qquad \qquad\qquad\qquad\qquad\qquad+ \big(\acal^{(N+1)}_{n[i+\binomaN\ j]} + \ccal^{(N+1)}_{n[ij]}\big)f^{N+1}_- \Big)\otimes \psi^{(N)}_{n-1[i]}\nonumber\\
&&+\sum_{i=1}^{\binomNnmdeux}\big(\ccal^{(N+1)}_{n[i+\binomNnmun\ j]}e^{-\alpha-(N-2n+4)\phi/2+i\theta}\big)f^{N+1}_+ +  \big(\ccal^{(N+1)}_{n[i+\binomNnmun\ j]}\big)f^{N+1}_-\Big)\otimes \psi^{(N)}_{n-2[i]}\ .
\label{eq3}
\eeqa
Identifying (\ref{eq1}) with (\ref{eq3}), the $N+1-th$ entries  are determined recursively in terms of the $N-th$ entries (see Appendix A) and the final result is consistent with the block tridiagonal structure. 
In particular,
\beqa
 \ccal^{(N+1)}_{n[i+\binomNmun \ j]}\equiv 0 \qquad\mbox{for}\qquad i\in \{1,...,\Big({{ N }\atop {n-2}}\Big)\}, \quad\ j\in \{1,...,\Big({{ N }\atop {n}}\Big)\}\ .\nonumber
\eeqa
Then,  (\ref{recpsi}) is satisfied for $N\rightarrow N+1$ and $j\in \{1,...,\big({{N}\atop {n}}\big)\}$, provided it holds for $N$. 
We now consider the action of ${\cal W}^{(N+1)}_1$ on the eigenvectors in the domain $j\in \{\big({{ N }\atop {n}}\big)+1,...,\big({{ N+1 }\atop {n}}\big)\}$. We follow the same approach as above: after a straightforward calculation, we find similarly that (\ref{recpsi}) for $N\rightarrow N+1$ is also satisfied, and in particular
\beqa
 \bcal^{(N+1)}_{n[ij]}\equiv 0 \qquad\mbox{for}\qquad i\in \{1,...,\Big({{ N }\atop {n+1}}\Big)\}, \quad\ j\in \{\Big({{ N }\atop {n}}\Big)+1,...,\Big({{ N +1}\atop {n}}\Big)\}\ .\nonumber
\eeqa
As a consequence, (\ref{recpsi}) holds for any $j\in \{\big(1,...,\big({{ N +1}\atop {n}}\big)$. We have checked  (\ref{recpsi}) for $N=1,2$ and $N=3$ - the case where some entries are vanishing - so we conclude that the assertion holds for all $N$. Explicit expressions for the entries are reported in Appendix A.\finproof

\begin{definition}
\label{}
Let $V$ denote the finite dimensional representation (of dimension $2^N$) on which ${\cal W}_{1}^{(N)}$ acts. Suppose ${\cal W}_{1}^{(N)}$ is diagonalizable on $V$, and let $V^*_0,V^*_1,V^*_2,...$ denote a standard ordering of the eigenspaces of ${\cal W}_{1}^{(N)}$.
For $s=0,1,2,...$, let ${\tilde\lambda}^{(N)}_{s}$ denote the eigenvalue sequence of ${\cal W}_{1}^{(N)}$.
\end{definition}

\begin{proposition}
\label{props}
Define $k_+=(k_-)^\dagger=-(q^{1/2}-q^{-1/2})e^{i\theta}/2$. Let $\{\alpha,\alpha^*\}\in{\mathbb C}^*$, $\theta\in{\mathbb R}$ be generic scalars and $\phi$ purely imaginary. Introduce the parameters ${\tilde \epsilon}^{[k]}_l=\pm 1$, $l=1,...,N$, $k\in\{1,...,\big({{ N }\atop {s}}\big)\}$. For any $N$, the set $\varphi^{(N)}_{s[k]}$ defined by:
\beqa
\varphi^{(N)}_{s[k]}=\bigotimes_{l=1}^N \left( -e^{-{\tilde \epsilon}^{[k]}_l\alpha^*-{\tilde \epsilon}^{[k]}_l\sum_{j=1}^{l-1}{\tilde \epsilon}^{[k]}_j\phi/2+i\theta}f^l_+ + f^l_- \right)\ ,\qquad s=(N-\sum_{j=1}^{N}{\tilde\epsilon}^{[k]}_j)/2\label{dualeigenvectN}
\eeqa
form a complete basis of $V$. In this basis, the matrices  ${\cal W}^{(N)}_0$, ${\cal W}^{(N)}_1$ are such that:
\beqa
{\cal W}^{(N)}_1\varphi^{(N)}_{s[k]}&=&{\tilde \lambda}^{(N)}_s\varphi^{(N)}_{s[k]} \qquad \mbox{with} \qquad {\tilde\lambda}^{(N)}_s=\cosh(\alpha^*+(N-2s)\phi/2)\ ,\nonumber\\
{\cal W}^{(N)}_0\varphi^{(N)}_{s[k]}&=& \sum_{l=1}^{\binomNspun}\bcalt^{(N)}_{s[lk]}\varphi^{(N)}_{s+1[l]} + \sum_{l=1}^{\binomNs}\acalt^{(N)}_{s[lk]}\varphi^{(N)}_{s[l]} + \sum_{l=1}^{\binomNsmun}\ccalt^{(N)}_{s[lk]}\varphi^{(N)}_{s-1[l]} \ ,\label{recvarphi}
\eeqa
where the coefficients $\acalt^{(N)}_{s[lk]},\bcalt^{(N)}_{s[lk]},\ccalt^{(N)}_{s[lk]}$ are obtained from Appendix A with the substitution (\ref{subs}). 
\label{diagW1}
\end{proposition}

{\it Proof}: According to the expressions (\ref{op}) and previous analysis, the proof is straightforward.\finproof

\begin{lemma}
\label{}
Let $\{\alpha,\alpha^*\}\in{\mathbb C}^*$, $\theta\in{\mathbb R}$ be generic scalars and $\phi$ purely imaginary. For any $N$, let the matrices ${\cal W}_{0}^{(N)},{\cal W}_{1}^{(N)}$ be defined by (\ref{op}).
Then, ${\cal W}_{0}^{(N)},{\cal W}_{1}^{(N)}$ act on $V$ as a TD pair of diameter $N$. 
\end{lemma}

{\it Proof}:  We show that the conditions ({\it 1-4}) of Definition \ref{defTDpair} are satisfied. According to Proposition \ref{diagW0} and Proposition \ref{diagW1}, the condition {\it 1} is satisfied.
Having $n\in \{0,1,...,N\}$ and the block tridiagonal structure (\ref{recpsi}), (\ref{recvarphi}),  $V=\bigoplus_{n=0}^N V_n$ is irreducible i.e. condition {\it 4} is satisfied. Finally, Proposition \ref{tridiagW1} and Proposition \ref{diagW1} imply that the conditions ({\it 2,3}) are satisfied. Then, ${\cal W}_{0}^{(N)},{\cal W}_{1}^{(N)}$ is a TD pair. From Corollary \ref{nombreval} and Proposition \ref{props} one finds that the diameter of the pair is $d=N$.\finproof  

\begin{remark}
\label{}
In agreement with \cite{Ter01}, the eigenvalue sequence $(\lambda_0^{(N)},\lambda_1^{(N)},...,\lambda_N^{(N)})$ and dual eigenvalue sequence $(\lambdat_0^{(N)},\lambdat_1^{(N)},...,\lambdat_N^{(N)})$ satisfy
\beqa
\frac{\lambda_{n-2}^{(N)}-\lambda_{n+1}^{(N)}}{\lambda_{n-1}^{(N)}-\lambda_{n}^{(N)}}= q+q^{-1}+1 \qquad \quad \mbox{and}\qquad \qquad \frac{\lambdat_{s-2}^{(N)}-\lambdat_{s+1}^{(N)}}{\lambdat_{s-1}^{(N)}-\lambdat_{s}^{(N)}}= q+q^{-1}+1\nonumber
\eeqa
for $n,s=2,...,N-1$. The parameters $\rho,\rho^*$ in (\ref{qDG}) are given by $\rho=\rho^*=-(q-q^{-1})^2/4$. 
\end{remark}

\section{Examples of related symmetric functions}
As suggested in \cite{Ter03}, given the connection between Leonard pairs\,\footnote{Leonard pairs are the simplest examples of tridiagonal pairs. Their shape vector being $(1,...,1)$, all eigenspaces have dimension one. In particular, in the basis that diagonalizes ${\textsf A}$ (resp. ${\textsf A}^*$), the matrix representing ${\textsf A}^*$ (resp. ${\textsf A}$) is irreducible tridiagonal.} and the Askey-Wilson orthogonal polynomials ($q-$Racah, $q-$Hahn,...) on a discrete support it is interesting to investigate the family of functions associated with more general TD pairs, which is the purpose of this Section.      
As shown in \cite{Zhed92}, Askey-Wilson polynomials of general form with a discrete argument satisfy a three term recurrence relation and a second-order $q-$difference equation that can be derived starting from the structure of the finite dimensional representations of the Askey-Wilson algebra. In particular, to the four parameters of the polynomials correspond the four independent structure constants of the algebra. Following a similar point of view, for the TD pairs (\ref{TDpairgen}) it is possible to deduce a coupled system of recurrence relations and a set of second-order $q-$difference equations. Below, we consider in particular the family (\ref{op}) for $\alpha,\alpha^*$ purely imaginary\,\footnote{Similar analysis can be done for $\alpha^\dagger=\alpha^*$ in which case $({\cal W}^{(N)}_0)^\dagger\equiv {\cal W}^{(N)}_1$.} so that $\{\lambda^{(N)}_n,\lambdat^{(N)}_s\}\in {\mathbb R}$. For this choice, $({\cal W}^{(N)}_0)^\dagger\equiv {\cal W}^{(N)}_1|_{\alpha^*\rightarrow \alpha}$ for any $N$. So, we define ${\tilde \varphi}^{(N)}_{s[k]}\equiv \psi^{(N)}_{n[k]}|_{\alpha\rightarrow \alpha^*,n\rightarrow s }$\ .  
\subsection{Example $N=1$}
For $N=1$, according to previous analysis the pair ${\cal W}_{0}^{(1)}$, ${\cal W}_{1}^{(1)}$ is a Leonard pair. By analogy with \cite{Zhed92},
using the scalar product in the canonical basis $f_\pm^{(1)}$ we introduce the  function ${\cal F}^{(1)}_{n[1]}(\lambdat)$ for $n=0,1$ - the index $k=1$ is ommited for simplicity - such that
\beqa
\langle {\tilde\varphi}^{(1)}_{s[1]}, \psi^{(1)}_{n[1]} \rangle = U^{(1)}_1(s) {\cal F}^{(1)}_{n[1]}(\lambdat^{(1)}_s)\qquad \mbox{with}\qquad {\cal F}^{(1)}_{0[1]}(\lambdat^{(1)}_s)\equiv 1 \label{defF1}
\eeqa
for all $s=0,1$. For this choice of normalization, the coefficient $U^{(1)}_1(s)=e^{\alpha-(1-2s)\alpha^*}+1$\ , as follows from the definitions (\ref{eigenvect1}). Using the matrix representation (\ref{init}) for ${\cal W}_{1}^{(1)}$, the expression $\langle {\tilde\varphi}^{(1)}_{s[1]}, {\cal W}_{1}^{(1)}\psi^{(1)}_{n[1]} \rangle $ yields the following (two term) recurrence relations:
\beqa
\lambdat^{(1)}_s {\cal F}^{(1)}_{0[1]}(\lambdat^{(1)}_s)&=&\bcal^{(1)}_{0}{\cal F}^{(1)}_{1[1]}(\lambdat^{(1)}_s) + \acal^{(1)}_{0}{\cal F}^{(1)}_{0[1]}(\lambdat^{(1)}_s)\ ,\nonumber\\
\lambdat^{(1)}_s {\cal F}^{(1)}_{1[1]}(\lambdat^{(1)}_s)&=&\acal^{(1)}_{1}{\cal F}^{(1)}_{0[1]}(\lambdat^{(1)}_s) + \ccal^{(1)}_{1}{\cal F}^{(1)}_{1[1]}(\lambdat^{(1)}_s)\ \label{recN1}
\eeqa
for any $s=0,1$. For the choice (\ref{defF1}), the first equation determines uniquely ${\cal F}^{(1)}_{1[1]}(\lambdat)=\big(\lambdat -  \acal^{(1)}_{0}\big)/\bcal^{(1)}_{0}$ which form coincides with the first of the Askey-Wilson polynomials with argument $\lambdat$. Replacing this latter expression in the second equation of (\ref{recN1}), the values of  $\lambdat$ are restricted to   $\lambdat=\cosh(\alpha^*\pm\phi/2)$, in perfect agreement with the structure of the eigenvalues in the dual basis as found in Section 3.1. \vspace{1mm}

On the other hand, we may expand the expression $\langle {\tilde\varphi}^{(1)}_{s[1]}, {\cal W}_{0}^{(1)}\psi^{(1)}_{n[1]} \rangle $ using the definition (\ref{defF1}). In this case, we now get the following  (first order) $q-$difference equations:
\beqa
\lambda^{(1)}_n {\cal F}^{(1)}_{n[1]}(\lambdat^{(1)}_0)&=&\bcalt^{(1)}_{0}\frac{U^{(1)}_1(1)}{U^{(1)}_1(0)}{\cal F}^{(1)}_{n[1]}(\lambdat^{(1)}_1) + \acalt^{(1)}_{0}{\cal F}^{(1)}_{n[1]}(\lambdat^{(1)}_0)\ ,\nonumber\\
\lambda^{(1)}_n {\cal F}^{(1)}_{n[1]}(\lambdat^{(1)}_1)&=&\acalt^{(1)}_{1}{\cal F}^{(1)}_{n[1]}(\lambdat^{(1)}_1) + \ccalt^{(1)}_{1}\frac{U^{(1)}_1(0)}{U^{(1)}_1(1)}{\cal F}^{(1)}_{n[1]}(\lambdat^{(1)}_0)\ \label{recN1}
\eeqa
for any $n=0,1$. The fact that we obtain a two term recurrence relation and a first order $q-$difference equation is not surprising, as we started with spin-$j=1/2$ representations of $U_{q^{1/2}}(sl_2)$. Indeed, three term recurrence relations and second-order $q-$difference equations appear for spin-$j>1/2$ representations of $U_{q^{1/2}}(sl_2)$ only.
\subsection{Example $N=2$}
For $N=2$, the pair ${\cal W}_{0}^{(2)}$, ${\cal W}_{1}^{(2)}$ is the first example of TD pair that is not a Leonard pair, due to the multiplicity of one of the eigenvalues. We introduce the fonction ${\cal F}^{(2)}_{n[ik]}(\lambdat)$ for any $n=0,1,2$ and $i\in \{\big(1,...,\big({{2}\atop {n}}\big)\}$ such that:
\beqa
\langle {\tilde\varphi}^{(2)}_{s[k]}, \psi^{(2)}_{n[i]} \rangle = U^{(2)}_k(s) {\cal F}^{(2)}_{n[ik]}(\lambdat^{(2)}_s) \label{defF2}
\eeqa
for $s=0,1,2$ and $k\in \{\big(1,...,\big({{2}\atop {s}}\big)\}$. Choosing the normalization 
\beqa
{\cal F}^{(2)}_{0[1k]}(\lambdat^{(2)}_s)\equiv 1 \qquad \mbox{for any}\qquad s=0,1,2,\label{normF2}
\eeqa
and using the scalar product in the canonical basis and the explicit form of the eigenvectors (\ref{eigenvect2}),
it follows
\beqa
U^{(2)}_1(s=0)&=& (e^{\alpha-\alpha^*}+1)^2\ ,\nonumber\\
U^{(2)}_1(s=1)&=& (e^{\alpha-\alpha^*+\phi}+1)(e^{\alpha+\alpha^*}+1)\ ,\qquad U^{(2)}_2(s=1)= (e^{\alpha+\alpha^*+\phi}+1)(e^{\alpha-\alpha^*}+1)\ ,\nonumber\\
U^{(2)}_1(s=2)&=& (e^{\alpha+\alpha^*}+1)^2\ .\nonumber
\eeqa
Then, using the matrix representation (\ref{mat2}) for ${\cal W}_{1}^{(2)}$ we immediately derive the following (coupled) system of recurrence relations
- the index $k$ is ommited for simplicity:
\beqa
\lambdat^{(2)}_s {\cal F}^{(2)}_{0[1]}(\lambdat^{(2)}_s)&=&\bcal^{(2)}_{0[11]} {\cal F}^{(2)}_{1[1]}(\lambdat^{(2)}_s) +  \bcal^{(2)}_{0[21]}{\cal F}^{(2)}_{1[2]}(\lambdat^{(2)}_s) + \acal^{(2)}_{0[11]} {\cal F}^{(2)}_{0[1]}(\lambdat^{(2)}_s) \label{i}\ ,\\
\lambdat^{(2)}_s {\cal F}^{(2)}_{1[1]}(\lambdat^{(2)}_s)&=&\bcal^{(2)}_{1[11]} {\cal F}^{(2)}_{2[1]}(\lambdat^{(2)}_s) +  \acal^{(2)}_{1[11]}{\cal F}^{(2)}_{1[1]}(\lambdat^{(2)}_s) + \acal^{(2)}_{1[21]} {\cal F}^{(2)}_{1[2]}(\lambdat^{(2)}_s) +\ccal^{(2)}_{1[11]} {\cal F}^{(2)}_{0[1]}(\lambdat^{(2)}_s) \label{ii}\ ,\\
\lambdat^{(2)}_s {\cal F}^{(2)}_{1[2]}(\lambdat^{(2)}_s)&=&\bcal^{(2)}_{1[12]} {\cal F}^{(2)}_{2[1]}(\lambdat^{(2)}_s) +  \acal^{(2)}_{1[12]}{\cal F}^{(2)}_{1[1]}(\lambdat^{(2)}_s) + \acal^{(2)}_{1[22]} {\cal F}^{(2)}_{1[2]}(\lambdat^{(2)}_s) +\ccal^{(2)}_{1[12]} {\cal F}^{(2)}_{0[1]}(\lambdat^{(2)}_s) \label{iii}\ ,\\
\lambdat^{(2)}_s {\cal F}^{(2)}_{2[1]}(\lambdat^{(2)}_s)&=& \acal^{(2)}_{2[11]}{\cal F}^{(2)}_{2[1]}(\lambdat^{(2)}_s) + \ccal^{(2)}_{2[11]} {\cal F}^{(2)}_{1[1]}(\lambdat^{(2)}_s) +\ccal^{(2)}_{2[21]} {\cal F}^{(2)}_{1[2]}(\lambdat^{(2)}_s) \ ,\label{iv}
\eeqa
for any $s=0,1,2$.
For the normalization (\ref{normF2}), a straightforward calculation shows that ${\cal F}^{(2)}_{n[i]}(\lambdat)$ are {\it rational} functions of $\lambdat$ uniquely determined by the equations (\ref{i}), (\ref{ii}) and (\ref{iv}) for any values of $\lambdat$. Explicitely, together with (\ref{normF2}) they read
\beqa
{\cal F}^{(2)}_{1[1]}(\lambdat)&=&  -\frac{e^{-\alpha-\phi/2}\sinh\phi(\cosh\alpha + \cosh\alpha^*)}{\sinh(\alpha-\phi/2)\cosh\alpha^* +\cosh(\alpha+\phi/2)\sinh\alpha-\sinh(\phi/2)}\ \frac{(\lambdat-u_{1[1]})}{(\lambdat-v)}
\ ,\nonumber \\
{\cal F}^{(2)}_{1[2]}(\lambdat)&=& \frac{e^{-\alpha-\phi/2}\sinh(\alpha+\phi/2)\sinh\alpha}{\sinh(\phi/2)(\sinh(\alpha-\phi/2)\cosh\alpha^* +\cosh(\alpha+\phi/2)\sinh\alpha-\sinh(\phi/2))}\ \frac{(\lambdat-u^{(+)}_{1[2]})(\lambdat-u^{(-)}_{1[2]})}{(\lambdat-v)}\ ,\nonumber \\
{\cal F}^{(2)}_{2[1]}(\lambdat)&=& -\frac{e^{-2\alpha}\sinh(\alpha+\phi/2)(\cosh\alpha + \cosh\alpha^*)}{\sinh(\alpha-\phi/2)\cosh\alpha^* +\cosh(\alpha+\phi/2)\sinh\alpha-\sinh(\phi/2)}\ \frac{(\lambdat-u_{2[1]})}{(\lambdat-v)}\ ,\label{sol2}
\label{solF2}
\eeqa
where the expressions $u_{1[1]},u^{(\pm)}_{1[2]},u_{2[1]}$ and $v$ are reported in Appendix B. Note that these expressions are invariant under the substitution $\alpha^*\rightarrow -\alpha^*$.
Finally, the equation (\ref{iii}) fixes the values of $\lambdat$: replacing  (\ref{solF2}) in (\ref{iii}) and using the explicit expressions of the coefficients, one gets a third order polynomial in $\lambdat$ which roots reproduce exactly the eigenvalue sequence (\ref{eigenvalN}) for $N=2$, as expected.\vspace{1mm}

On the other hand, it is possible to expand $\langle {\tilde\varphi}^{(2)}_{s[k]}, {\cal W}_{0}^{(2)}\psi^{(2)}_{n[i]} \rangle $ using (\ref{defF2}) which leads to a set of $q-$difference equations.
This being analogous to the one for higher values of $N$, we refer the reader to the analysis below. 
\subsection{Generalized recurrence relations and $q-$difference equations}
Let us now turn to general values of $N$. Due to the block tridiagonal structure of the matrices ${\cal W}_{0}^{(N)}$, ${\cal W}_{1}^{(N)}$ with respect to their dual basis, for arbirary $N$ it is easy to derive a system of recurrence relations and $q-$difference equations. By analogy with (\ref{defF2}), we introduce the fonctions ${\cal F}^{(N)}_{n[ik]}(\lambdat)$ for any $n=0,1,...,N$ and $i\in \{\big(1,...,\big({{N}\atop {n}}\big)\}$:
\beqa
\langle {\tilde\varphi}^{(N)}_{s[k]}, \psi^{(N)}_{n[i]} \rangle = U^{(N)}_k(s) {\cal F}^{(N)}_{n[ik]}(\lambdat^{(N)}_s) \qquad \mbox{with}\qquad {\cal F}^{(N)}_{0[1k]}(\lambdat^{(N)}_s)\equiv 1 \label{defFN}
\eeqa
for any $s=0,1,...,N$ and $k\in \{\big(1,...,\big({{N}\atop {s}}\big)\}$. Here, the coefficients $U^{(N)}_k(s)$ can be calculated recursively based on previous results: using the scalar product in the canonical basis together with the explicit form of the eigenvectors (\ref{eigenvectN}) for $s=0,1,...,N$ we find:
\beqa
U^{(N)}_k(s)&=& (e^{(\alpha-\alpha^*+s\phi)}+1)U^{(N-1)}_k(s)\qquad \qquad\qquad\ \mbox{for}\qquad k\in\{1,...,\Big({{N-1}\atop {s}}\Big)\}\ ,\nonumber\\
U^{(N)}_k(s)&=& (e^{(\alpha+\alpha^*+(N-s)\phi)}+1)U^{(N-1)}_{k-(\!{{N-1}\atop {s}}\!)}(s-1)\qquad \mbox{for}\qquad k\in\{\Big({{N-1}\atop {s}}\Big)+1,...,\Big({{N}\atop {s}}\Big)\}\ .\label{Uks} 
\eeqa 
First, let us focus on the system of recurrence relations for the rational fonctions ${\cal F}^{(N)}_{n[ik]}(\lambdat)$. Using the results of previous Sections, from $\langle{\tilde\varphi}^{(N)}_{s[k]}, {\cal W}_{1}^{(N)}\psi^{(N)}_{n[i]} \rangle$ and ommiting the index $k$ we immediatly get 
\beqa
\lambdat {\cal F}^{(N)}_{n[j]}(\lambdat)&=& \sum_{i=1}^{\binombN}\bcal^{(N)}_{n[ij]}{\cal F}^{(N)}_{n+1[i]}(\lambdat) + \sum_{i=1}^{\binomaN}\acal^{(N)}_{n[ij]}{\cal F}^{(N)}_{n[i]}(\lambdat) + \sum_{i=1}^{\binomcN}\ccal^{(N)}_{n[ij]}{\cal F}^{(N)}_{n-1[i]}(\lambdat)\label{recFN}
\eeqa
for $\lambdat\in\{\lambdat^{(N)}_s\}$,  $n,s=0,1,...,N$, $j\in\{1,...,\binomN\}$ with the coefficients given in Appendix A. Notice that similarly to the case $N=2$, $2^N-1$ equations will determine uniquely the explicit form of ${\cal F}^{(N)}_{n[i]}(\lambdat)$. The remaining equation, associated with  $n=N-1$ and $j=N$, determines the eigenvalues to be in the set $\lambdat\in\{\lambdat^{(N)}_s\}$,  $s=0,1,...,N$ given by (\ref{recvarphi}).\vspace{1mm}

To derive a system of $q-$difference equations for the rational fonctions ${\cal F}^{(N)}_{n[ik]}(\lambdat_s)$, we start from $\langle{\tilde\varphi}^{(N)}_{s[k]}, {\cal W}_{0}^{(N)}\psi^{(N)}_{n[i]} \rangle$. Let us introduce the discrete $q-$difference operator $\eta^{\pm 1}g(\lambdat_s)=g(\lambdat_{s\pm 1})$. Combining the normalization coefficients $U^{(N)}_k(s)$ given by (\ref{Uks}) with the ``dual'' entries we obtain a degenerate family of second-order $q-$difference operators such that
\beqa
{{\mathbb D}}^{(N)}(s){\mathbb F}^{(N)}_{n[j]}(\lambdat^{(N)}_s)=\lambda^{(N)}_n {\mathbb F}^{(N)}_{n[j]}(\lambdat^{(N)}_s)\ \qquad \mbox{with}\qquad {\mathbb F}^{(N)}_{n[j]} = \big[{\cal F}^{(N)}_{n[j1]},{\cal F}^{(N)}_{n[j2]},...,{\cal F}^{(N)}_{n[j({{N}\atop {s}})]}\big]^t \label{qdiffFN}
\eeqa
where
\beqa
\qquad{{\mathbb D}}^{(N)}(s)\equiv \Phi^{(N)}(s)\eta + \overline{\Phi}^{(N)}(s)\eta^{-1} + \mu^{(N)}(s)\ ,\nonumber
\eeqa
\beqa 
\big[\Phi^{(N)}(s)\big]_{kl}&=&\bcalt^{(N)}_{s[lk]}\frac{U^{(N)}_l(s+1)}{U^{(N)}_k(s)}\ ,\qquad \big[\overline{\Phi}^{(N)}(s)\big]_{kl}=\ccalt^{(N)}_{s[lk]}\frac{U^{(N)}_l(s-1)}{U^{(N)}_k(s)}\ ,\nonumber\\
\big[\mu^{(N)}(s)\big]_{kl}&=&\acalt^{(N)}_{s[lk]}\frac{U^{(N)}_l(s)}{U^{(N)}_k(s)}\ .\nonumber
\eeqa
It should be pointed out that for $N=1$, the operator ${\mathbb D}^{(1)}(s)$ is a special case (as we choosed the spin$-1/2$ representation of $U_{q^{1/2}}(sl_2)$) of the Askey-Wilson operator derived in \cite{Zhed92}. As a consequence, for higher values of $N$ the $q-$difference operators ${\mathbb D}^{(N)}(s)$ can be seen as some generalizations of the Askey-Wilson operator. Although it is an interesting problem to explore some consequences in integrable systems and the links with a recent work on Bethe ansatz and $q-$Sturm-Liouville problems \cite{Ismail}, we do not pursue the analysis further and leave it for a separate work \cite{prep}. Note that for $N=2$ we have checked explicitly that the rational functions (\ref{sol2}) satisfy the degenerate set of $q-$difference equations (\ref{qdiffFN}). 
 
\subsection{Orthogonality}
The system of Askey-Wilson polynomials is orthogonal with respect to a certain discrete weight function. This can be shown, for instance, using the completness of the two dual basis which diagonalize the elements of the Askey-Wilson algebra \cite{Zhed92}. Such property can be extended to the system of rational functions ${\cal F}^{(N)}_{n[jk]}(\lambdat)$ (defined on a real discrete support) obeying (\ref{recFN}), (\ref{qdiffFN}), having in mind that the two basis (\ref{eigenvectN}), (\ref{dualeigenvectN}) are complete for general values of the parameters. To show that, we introduce the basis of dual eigenvectors ${\tilde \varphi}^{(N)}_{s[k]}\equiv \psi^{(N)}_{n[k]}|_{\alpha\rightarrow \alpha^*,n\rightarrow s }$ and ${\tilde \psi}^{(N)}_{n[i]}\equiv \varphi^{(N)}_{n[i]}|_{\alpha^*\rightarrow \alpha,s\rightarrow n }$. It is easy to check that
\beqa
{\cal N}^{(N)}_{n[i]}\langle{\tilde \psi}^{(N)}_{n[i]},\psi^{(N)}_{m[j]} \rangle = \delta_{n,m}\delta_{i,j}\qquad \mbox{and}\qquad {\tilde {\cal N}}^{(N)}_{s[k]}\langle{\tilde\varphi}^{(N)}_{s[k]},\varphi^{(N)}_{r[l]} \rangle = \delta_{s,r}\delta_{k,l} \ ,
\eeqa
where, using the explicit expression of the eigenvectors, one can easily obtain recursively the normalization coefficients. For instance, for $n=0,...,N$ one finds
\beqa
{\cal N}^{(N)}_{n[i]}&=& {\cal N}^{(N-1)}_{n[i]}\big(1-e^{(2\alpha+(N-1-2n)\phi)}\big)^{-1}\qquad \qquad\qquad\ \mbox{for}\qquad i\in\{1,...,\Big({{N-1}\atop {n}}\Big)\}\ ,\nonumber\\
{\cal N}^{(N)}_{n[i]}&=& {\cal N}^{(N-1)}_{n-1[i-(\!{{N-1}\atop {n}}\!)]}\big(1-e^{(-2\alpha-(N+1-2n)\phi)}\big)^{-1}\qquad \ \qquad \mbox{for} \qquad i\in\{\Big({{N-1}\atop {n}}\Big)+1,...,\Big({{N}\atop {n}}\Big)\}\ ,
\label{normbasis}
\eeqa
and similarly ${\tilde {\cal N}}^{(N)}_{s[k]}={\cal N}^{(N)}_{n[i]}|_{\{n,i,\alpha,\phi\}\rightarrow\{s,k,-\alpha^*,-\phi\}}$\  for $s=0,1,...,N$.   
Then, the completness of the two eigenbasis together with the definition of the normalization coefficients yield
\beqa
\langle{\tilde\psi}^{(N)}_{n[i]},\psi^{(N)}_{m[j]} \rangle &=&   \sum_{s=0}^N \sum_{k=1}^{\big({{N}\atop {s}}\big)}{\tilde {\cal N}}^{(N)}_{s[k]}\langle{\tilde\psi}^{(N)}_{n[i]},\varphi^{(N)}_{s[k]}\rangle \langle{\tilde\varphi}^{(N)}_{s[k]},\psi^{(N)}_{m[j]} \rangle\ .\nonumber
\eeqa
Replacing (\ref{defFN}) in the above equation, and using $\langle{\tilde\psi}^{(N)}_{n[i]},\varphi^{(N)}_{s[k]}\rangle =\langle{\tilde\varphi}^{(N)}_{s[k]},\psi^{(N)}_{m[j]} \rangle$ we find that the system of rational functions ${\cal F}^{(N)}_{n[jk]}(\lambdat_s)$ of the discrete argument $\lambdat_s$ are orthogonal on the $N+1$ points on the interval of the real axis $0 \leq s \leq N$. The condition of orthogonality reads 
\beqa
\sum_{s=0}^N \sum_{k=1}^{\big({{N}\atop {s}}\big)} w_k^{(N)}(s)  {\cal F}^{(N)}_{n[ik]}(\lambdat_s) {\cal F}^{(N)}_{m[jk]}(\lambdat_s) = \big({{\cal N}^{(N)}_{n[i]}}\big)^{-1}\delta_{n,m}\delta_{i,j}\qquad \mbox{where}\qquad
w_k^{(N)}(s) ={\tilde {\cal N}}^{(N)}_{s[k]} \big({U^{(N)}_{k}(s)}\big)^2 \nonumber 
\eeqa
is the discrete weight function that ensures the orthogonality of the sytem. Given $N$, its explicit expression is derived recursively using  (\ref{normbasis}), (\ref{Uks}). In particular, it is an exercise to check the orthogonality condition for $N=1$ (in which case our construction reduces to a special case (spin$-1/2$) of \cite{Zhed92}) and $N=2$. 
\vspace{4mm}

\noindent{\bf Acknowledgements:} I thank  H. Giacomini, K. Koizumi and K. Noui for important discussions, and P. Terwilliger for suggestions and interest in this work. I wish to thank the organizers of the 3rd Annual EUCLID Meeting 2005 where preliminary results were presented. Part of this work is supported by the ANR research project ``{\it Boundary integrable models: algebraic structures and correlation functions}'', contract number JC05-52749 and TMR Network EUCLID ``{\it Integrable models and applications: from strings to condensed matter}'', contract number HPRN-CT-2002-00325.\vspace{0.5cm}

\vspace{5mm}

\centerline{\bf \large Appendix A: Coefficients of the matrices $A^{(N)}_n$, $B^{(N)}_n$, $C^{(N)}_n$}
\vspace{4mm}

We can organize the set of eigenvectors by eigenspaces $V_n$ and set $\Psi^{(N)}_n = \big[\psi^{(N)}_{n[1]},\psi^{(N)}_{n[2]},...,\psi^{(N)}_{n[\binomaN]}\big]^t$. Then,
the relations (\ref{specvalN}), (\ref{recpsi}) can be written in the form
\beqa
{\cal W}^{(N)}_0\Psi^{(N)}_n &=& \lambda_n^{(N)} \Psi^{(N)}_n\ ,\nonumber\\
{\cal W}^{(N)}_1\Psi^{(N)}_n &=& B^{(N)}_n\Psi^{(N)}_{n+1} + A^{(N)}_n\Psi^{(N)}_{n} + C^{(N)}_n\Psi^{(N)}_{n-1}\ ,
\eeqa
where $A^{(N)}_n,B^{(N)}_n,C^{(N)}_n$ are submatrices with entries $(X^{(N)}_n)_{[ji]}\equiv (\xcal^{(N)}_n)_{[ij]}$ which have been determined recursively in Section 3, with ``initial'' conditions (\ref{init}). The size of each submatrix is determined by the multiplicity of the eigenvalue. We obtain:\vspace{4mm}\\
$\bullet$ {\bf Matrix $A^{(N)}_n$}
\beqa
&&\mbox{For} \quad i\in\{1,...,\binomNmun\}:\qquad \acal^{(N)}_{n[ii]}=\frac{\acal^{(N-1)}_{n[ii]}\sinh(\alpha+(N-2-2n)\phi/2)-\sinh(\phi/2)}{\sinh(\alpha+(N-1-2n)\phi/2)}\ ,\nonumber\\ 
&&\mbox{For} \quad i\in\{\binomNmun+1,...,\binomN\}:\qquad
 \acal^{(N)}_{n[ii]}=\frac{\acal^{(N-1)}_{n-1[i-\binomNmun\ i-\binomNmun]}\sinh(\alpha+(N+2-2n)\phi/2)+\sinh(\phi/2)}{\sinh(\alpha+(N+1-2n)\phi/2)}\ ,\nonumber\\
&&\mbox{For} \quad  i\in\{1,...,\binomNmun\},\ j\in\{\binomNmun+1,...,\binomN\}:\qquad \acal^{(N)}_{n[ij]}=-\frac{e^{-\alpha-(N-2n)\phi/2}\sinh(\phi)}{\sinh(\alpha+(N-1-2n)\phi/2)}\bcal^{(N-1)}_{n-1[ij-\binomNmun]}\ ,\nonumber\\
&&\mbox{For} \quad i\in\{\binomNmun+1,...,\binomN\}, \quad j\in\{1,...,\binomNmun\}:\qquad \acal^{(N)}_{n[ij]}=\frac{e^{\alpha+(N-2n)\phi/2}\sinh(\phi)}{\sinh(\alpha+(N+1-2n)\phi/2)}\ccal^{(N-1)}_{n[i-\binomNmun\ j]}\ ,\nonumber\\
&&\mbox{For} \quad i,j\in\{1,...,\binomNmun\}, \quad i\neq j\ :\qquad \acal^{(N)}_{n[ij]}=\frac{\sinh(\alpha+(N-2-2n)\phi/2)}{\sinh(\alpha+(N-1-2n)\phi/2)}\acal^{(N-1)}_{n[ij]}\ ,\nonumber\\
&&\mbox{For} \quad i,j\in\{\binomNmun+1,...,\binomN\}, \quad i\neq j:\qquad \acal^{(N)}_{n[ij]}=\frac{\sinh(\alpha+(N+2-2n)\phi/2)}{\sinh(\alpha+(N+1-2n)\phi/2)}\acal^{(N-1)}_{n-1[i-\binomNmun\ j-\binomNmun]}\ .\nonumber
\eeqa
\vspace{4mm}
$\bullet$ {\bf Matrix $B^{(N)}_n$}
\beqa
&&\mbox{For} \quad i\in\{1,...,\binomNmunpun\}, \quad j\in\{1,...,\binomNmun\}:\qquad \bcal^{(N)}_{n[ij]}=e^{\phi/2}\bcal^{(N-1)}_{n[ij]}\ ,\nonumber \\
&&\mbox{For} \quad i\in\{1,...,\binomNmunpun\}, \quad j\in\{\binomNmun+1,...,\binomN\}:\qquad \bcal^{(N)}_{n[ij]}=0\ , \nonumber \\
&&\mbox{For} \quad i\in\{\binomNmunpun+1,...,\binomNpun\}, \quad j\in\{\binomNmun+1,...,\binomN\}:\nonumber\\ &&\qquad \qquad\qquad\qquad \qquad \qquad \bcal^{(N)}_{n[ij]}=e^{-\phi/2}\frac{\sinh(\alpha+(N+1-2n)\phi/2)}{\sinh(\alpha+(N-1-2n)\phi/2)}\bcal^{(N-1)}_{n-1[i-\binomNmunpun\ j-\binomNmun]}\ ,
\nonumber\\
&&\mbox{For} \quad i\in\{\binomNmunpun+1,...,\binomNpun\}:\nonumber \\
&&\qquad\qquad \qquad \bcal^{(N)}_{n[i\ i-\binomNmunpun]}=\frac{e^{\alpha+(N-1-2n)\phi/2}\sinh(\phi/2)}{\sinh(\alpha+(N-1-2n)\phi/2)}\Big(  \acal^{(N-1)}_{n[i-\binomNmunpun\ i-\binomNmunpun]} + \cosh(\alpha+(N-1-2n)\phi/2)\Big)\ , \nonumber\\ 
&& \mbox{For} \quad i\in\{\binomNmunpun+1,...,\binomNpun\}, j\in\{1,...,\binomNmun\},\quad i\neq j+\binomNmunpun:\nonumber\\
&&\qquad \qquad\qquad\qquad \qquad\qquad\bcal^{(N)}_{n[ij]}=\frac{e^{\alpha+(N-1-2n)\phi/2}\sinh(\phi/2)}{\sinh(\alpha+(N-1-2n)\phi/2)}\acal^{(N-1)}_{n[i-\binomNmunpun\ j]}\ . \nonumber 
\eeqa
\vspace{4mm}
$\bullet$ {\bf Matrix $C^{(N)}_n$}
\beqa
&&\mbox{For} \quad i\in\{\binomNmunmun+1,...,\binomNnmun\}, \quad j\in\{\binomNmun+1,...,\binomN\}:\qquad \ccal^{(N)}_{n[ij]}=e^{\phi/2}\ccal^{(N-1)}_{n-1[i-\binomNmunmun\ j-\binomNmun]}\qquad \nonumber \\
&&\mbox{For} \quad i\in\{\binomNmunmun+1,...,\binomNnmun\}, \quad j\in\{1,...,\binomNmun\}:\qquad\qquad\ \ \ccal^{(N)}_{n[ij]}=0\ ,\nonumber \\
&&\mbox{For} \quad i\in\{1,...,\binomNmunmun\}\ , \quad j\in\{1,...,\binomNmun\}:\qquad \ccal^{(N)}_{n[ij]}=e^{-\phi/2}\frac{\sinh(\alpha+(N-1-2n)\phi/2)}{\sinh(\alpha+(N+1-2n)\phi/2)}\ccal^{(N-1)}_{n[ij]}\qquad\ ,\nonumber\\
&&\mbox{For} \quad i\in\{1,...,\binomNmunmun\}:\nonumber\\
&&\qquad\qquad\qquad \ccal^{(N)}_{n[i\ i+\binomNmun]}=-\frac{e^{-\alpha-(N+1-2n)\phi/2}\sinh(\phi/2)}{\sinh(\alpha+(N+1-2n)\phi/2)}\Big(\acal^{(N-1)}_{n-1[ii]} + \cosh(\alpha+(N+1-2n)\phi/2)\Big)\ , \nonumber\\ 
&&\qquad\qquad\qquad\qquad\qquad\qquad\qquad \ ,\nonumber \\
&&\mbox{For} \quad i\in\{1,...,\binomNmunmun\}, j\in\{\binomNmun+1,...,\binomN\},\quad i\neq j-\binomNmun:\nonumber\\ &&\qquad\qquad\qquad\ccal^{(N)}_{n[ij]}=-\frac{e^{-\alpha-(N+1-2n)\phi/2}\sinh(\phi/2)}{\sinh(\alpha+(N+1-2n)\phi/2)}\acal^{(N-1)}_{n-1[ij-\binomNmun]}\ .\nonumber
\eeqa

\vspace{1cm}

\centerline{\bf \large Appendix B: Zeroes and poles of the rational functions for $N=2$}
\vspace{-1mm}
\beqa
u_{1[1]}&=& -\cosh\alpha\ ,\nonumber \\
v&=&\!\!\! \ \frac{\sinh(\alpha+2\alpha^*+\phi/2)+\sinh(\alpha-2\alpha^*+\phi/2)+\sinh(2\alpha+\alpha^*+3\phi/2)+\sinh(2\alpha-\alpha^*+3\phi/2)}{2(\sinh(\alpha-\alpha^*-\phi/2)+\sinh(\alpha+\alpha^*-\phi/2)+\sinh(2\alpha+\phi/2)-3\sinh(\phi/2))}\nonumber\\
&+&\!\!\!\frac{\sinh(\alpha+\phi/2)-3\sinh(\alpha-\phi/2)+\sinh(\alpha-3\phi/2)-3\sinh(\alpha+3\phi/2)+3\sinh(\alpha^*+\phi/2)-3\sinh(\alpha^*-\phi/2)}{2(\sinh(\alpha-\alpha^*-\phi/2)+\sinh(\alpha+\alpha^*-\phi/2)+\sinh(2\alpha+\phi/2)-3\sinh(\phi/2))}
,\nonumber \\
u^{(\pm)}_{1[2]}&=&\frac{U \pm 2\sinh(\phi/2)\sqrt{V}}{4(\cosh(\phi/2)-\cosh(2\alpha+\phi/2))}\ ,\nonumber\\
u_{2[1]}&=& \frac{\sinh(\alpha-\alpha^*-\phi/2)+\sinh(\alpha+\alpha^*-\phi/2)-\sinh(\phi/2)-\sinh(3\phi/2)}{2\sinh(\alpha+\phi/2)}\ ,\nonumber  
\eeqa
where
\beqa
U&=&4\cosh(\alpha+\phi/2)+2\cosh(\alpha^*+\phi/2)+2\cosh(\alpha^*-\phi/2)-2\cosh(\alpha+3\phi/2)-2\cosh(\alpha-\phi/2)\nonumber\\
&&-\cosh(\alpha^*+2\alpha-\phi/2)-\cosh(\alpha^*-2\alpha+\phi/2)-\cosh(\alpha^*+2\alpha+3\phi/2)-\cosh(\alpha^*-2\alpha-3\phi/2)\ ,\nonumber\\
V&=&(\cosh(2\alpha)+\cosh(2\alpha+\phi)-2)\times\nonumber\\
&&\times\big(8+4\cosh(\alpha+\alpha^*)+4\cosh(\alpha-\alpha^*)+4\cosh(\alpha+\alpha^*+\phi)+4\cosh(\alpha-\alpha^*+\phi)+4\cosh(\phi)-2\cosh(2\alpha)\ \nonumber\\
&&+2\cosh(2\alpha+\phi)+\cosh(2\alpha+2\alpha^*)+\cosh(2\alpha-2\alpha^*)+\cosh(2\alpha+2\alpha^*+\phi)+\cosh(2\alpha-2\alpha^*+\phi)\big)\ .\nonumber
\eeqa

\end{document}